\documentclass[aps,prd,onecolumn,showpacs,showkeys,amsmath,amssymb]{revtex4}
\usepackage{epsfig} 
\usepackage{amsmath}  
\usepackage{makecell}
\usepackage{graphicx}
\usepackage{array}
\usepackage{subfigure}
\usepackage{bm}
\usepackage{longtable}
\usepackage{tabularx}
\usepackage{booktabs}
\usepackage{appendix}
\usepackage{caption2}
\usepackage{amsfonts}
\usepackage{amsmath}
\usepackage{graphicx}
\usepackage{subfigure}
\usepackage{dcolumn}
\usepackage{siunitx}
\usepackage{bm}
\usepackage{booktabs}
\usepackage[utf8]{inputenc}

\graphicspath{{figs/}}
\usepackage{float}
\usepackage{longtable,lscape}
\usepackage{txfonts}
\usepackage{overpic}
\usepackage{amssymb}
\usepackage{indentfirst}
\usepackage{epsfig}
\usepackage{feynmf}   
\usepackage{epstopdf}   
\usepackage{slashed}  
\usepackage{multirow}

\usepackage[colorlinks, citecolor=blue,anchorcolor=red,menucolor=red, linkcolor=red,filecolor=red,runcolor=red,urlcolor=blue,frenchlinks=red]{hyperref}

\usepackage{ulem}
\usepackage[usenames,dvipsnames]{color}

\makeatletter
\newcommand{\figcaption}{\def\@captype{figure}\caption}
\newcommand{\tabcaption}{\def\@captype{table}\caption}

\newcommand{\Rmnum}[1]{\expandafter\@slowromancap\romannumeral #1@}
\def\hlinewd#1{%
  \noalign{\ifnum0=`}\fi\hrule \@height #1 \futurelet
   \reserved@a\@xhline}
\makeatother

\begin{document}
\title{$D$-wave excited $cs\bar{c}\bar{s}$ tetraquark states with $J^{PC}=1^{++}$ and $1^{+-}$}

\author{Zi-Yan Yang$^1$}
\author{Wei Chen$^1$}
\email{chenwei29@mail.sysu.edu.cn}
\affiliation{$^1$School of Physics, Sun Yat-Sen University, Guangzhou 510275, China}

\begin{abstract}
We study the mass spectra of D-wave excited $cs\bar{c}\bar{s}$ tetraquark states with $J^{PC}=1^{++}$ and $1^{+-}$ in both symmetric $\mathbf{6}_{cs}\otimes\bar{\mathbf{6}}_{\bar{c}\bar{s}}$ and antisymmetric $\bar{\mathbf{3}}_{cs}\otimes\mathbf{3}_{\bar{c}\bar{s}}$ color configurations by using the QCD sum rule method. We construct the D-wave diquark-antidiquark type of $cs\bar{c}\bar{s}$ tetraquark interpolating currents in various excitation structures with $(L_\lambda,L_\rho\{l_{\rho_1},l_{\rho_2}\})=(2,0\{0,0\}),(1,1\{1,0\}),(1,1\{0,1\}),(0,2\{1,1\}),(0,2\{2,0\}),(0,2\{0,2\})$. Our results support to interpret the recently observed $X(4685)$ resonance as a D-wave $cs\bar{c}\bar{s}$ tetraquark state with $J^{PC}=1^{++}$ in the $(2,0\{0,0\})$ or $(0,2\{2,0\})$ excitation mode, although some other possible excitation structures cannot be excluded exhaustively within theoretical errors. 
Moreover, our results provide a mass relation $6_{\rho\rho}<3_{\lambda\lambda}<3_{\lambda\rho}<3_{\rho\rho}$ and $6_{\rho\rho}<3_{\lambda\lambda}<6_{\lambda\lambda}<3_{\rho\rho}$ for the positive and negative $\mathbb{C}$-parity D-wave $cs\bar{c}\bar{s}$ tetraquarks, respectively. We suggest searching for these possible D-wave $cs\bar{c}\bar{s}$ tetraquarks in both the hidden-charm channels $J/\psi\phi$, $\eta_c\phi$ and open-charm channels such as $D_s\bar{D}_s^*$, $D_{s}\bar{D}_{s1}^*$ and so on. 
\end{abstract}

\pacs{12.39.Mk, 12.38.Lg, 14.40.Ev, 14.40.Rt}
\keywords{Tetraquark states, exotic states, QCD sum rules}
\maketitle

\section{Introduction}\label{Sec:Intro}
The history of multiquarks has lasted over half a century after Gell-mann and Zweig proposing such configurations in building the quark model in 1964~\cite{GellMann1964,1964-Zweig-p-}. However, the existence of tetraquark and pentaquark has long been speculated but rarely, if ever, proven. The story has been changed since 2003 when lots of charmoniumlike/bottomoniumlike XYZ states~\cite{ParticleDataGroup:2022pth}, hidden-charm $P_c$ states~\cite{LHCb:2015yax,LHCb:2019kea,LHCb:2020jpq,LHCb:2021chn}, doubly-charm $T_{cc}^+$~\cite{LHCb:2021vvq,LHCb:2021auc} and fully-charm tetraquark~\cite{LHCb:2020bwg} states have been observed, which can not be well explained within traditional quark model. They are very good candidates of tetraquark and pentaquark states. One can find the experimental and theoretical progresses in some review papers~\cite{Chen:2016qju,Esposito:2016noz,Guo:2017jvc,Liu:2019zoy,Brambilla:2019esw,Chen:2022asf,Meng:2022ozq}.

In 2017, the LHCb Collaboration observed four $J/\psi\phi$ structures $X(4140)$, $X(4274)$, $X(4500)$ and $X(4700)$ in the $B^+\to J/\psi\phi K^+$ decay process~\cite{LHCb:2016nsl,LHCb:2016axx}, in which the $X(4140)$ and $X(4274)$ were confirmed in consistent with previous measurements by the CDF Collaboration~\cite{CDF:2009jgo,CDF:2011pep}, CMS Collaboration~\cite{CMS:2013jru}, D0 Collaboration~\cite{D0:2013jvp} and BABAR Collaboration~\cite{BaBar:2014wwp}, while the $X(4500)$ and $X(4700)$ were new resonances. Inspired by these structures observed in the $J/\psi\phi$ invariant mass spectrum, the $X(4140)$ and $X(4274)$ were considered to be the $cs\bar c\bar s$ tetraquark ground states while the $X(4500)$ and $X(4700)$ were interpreted as the $cs\bar c\bar s$ tetraquark excited states in various theoretical methods~\cite{Ebert2008,Chen:2010ze,Stancu2009,Chen2017a,Lu2016,Wu2016,Liu2021,Deng2017,Wang2016b}. 

Recently, the LHCb Collaboration performed an improved full amplitude analysis of the $B^+\to J/\psi\phi K^+$ decay using 6 times larger signal yield than previously analyzed~\cite{LHCb:2021uow}. They confirmed the four previously reported $J/\psi\phi$ states in Refs.~\cite{LHCb:2016nsl,LHCb:2016axx}. In addition, a new $X(4685)$ state was also observed in the $J/\psi\phi$ final state with $15\sigma$ significance, while its spin-parity was determined to be $J^P=1^+$.  Considering its observed channel, the quantum numbers of $X(4685)$ should be $J^{PC}=1^{++}$ with positive charge conjugation parity. Its mass and decay width are measured as $m=4684\pm7^{+13}_{-16}$ MeV and $\Gamma=126\pm15^{+37}_{-41}$ MeV. One may wonder if the $X(4685)$ and $X(4700)$ be the same resonance since they were observed in the same final states with very similar masses and decay widths. However, LHCb determined their spin-parity as $J^P=1^+$ for $X(4685)$ and $J^P=0^+$ for $X(4700)$~\cite{LHCb:2021uow}. They are definitely two different states. 

After the observations of the above $J/\psi\phi$ resonances, there have been some efforts to understand their underlying structures in the diquark-antidiquark picture. If the $X(4140)$ and $X(4274)$ can be assigned as the S-wave $cs\bar c\bar s$ tetraquark ground states with $J^{PC}=1^{++}$, the $X(4685)$ state may be interpreted as the D-wave $cs\bar c\bar s$ excited tetraquark state. In Ref.~\cite{Ebert2008}, the authors calculated the masses of the excited hidden-charm tetraquarks without internal diquark excitation ($\lambda$-mode excitation) by using the relativistic quark model. The mass of the D-wave $cs\bar{c}\bar{s}$ tetraquark with $J^{PC}=1^{++}$ was calculated to be around 4.8 GeV. The same $\lambda$-mode excited $1^{++}$ D-wave $cs\bar{c}\bar{s}$ tetraquarks were also studied in the color flux-tube model with the masses around 4.9 GeV and 5.2 GeV for the $\bar{\mathbf{3}}_{cs}\otimes\mathbf{3}_{\bar{c}\bar{s}}$ and $\mathbf{6}_{cs}\otimes\bar{\mathbf{6}}_{\bar{c}\bar{s}}$ color structures, respectively~\cite{Deng:2019dbg}. In Ref.~\cite{Lu2016}, the D-wave $cs\bar{c}\bar{s}$ tetraquarks were investigated in different excitation modes by considering internal excited diquarks ($\rho$-mode excitation) in the relativistic quark model. The masses of the $\rho$-mode D-wave $cs\bar{c}\bar{s}$ tetraquarks with $J^{PC}=1^{++}$ were predicted as 4.6-4.7 GeV, which are much lower than the $\lambda$-mode tetraquarks. Besides, the authors in Ref.~\cite{Turkan2021} calculated the mass of ground state of $1^{++}$ S-wave $cs\bar{c}\bar{s}$ tetraquark to be about 4.6 GeV in QCD sum rules, which is much higher than those obtained in Ref.~\cite{Chen:2010ze}. In Ref.~\cite{Wang:2021ghk}, the $X(4685)$ was also considered as the axialvector $2S$ radial excited $cs\bar{c}\bar{s}$ tetraquark state. 

According to the above analyses and theoretical investigations, the newly observed $X(4685)$ state may be explained as a $\rho$-mode excited $D-$wave $cs\bar{c}\bar{s}$ tetraquark with $J^{PC}=1^{++}$. In this work, we shall systematically study the mass spectra of the D-wave $cs\bar{c}\bar{s}$ with $J^{PC}=1^{++}$ and $1^{+-}$ in both color symmetric $\mathbf{6}_{cs}\otimes\bar{\mathbf{6}}_{\bar{c}\bar{s}}$ and antisymmetric $\bar{\mathbf{3}}_{cs}\otimes\mathbf{3}_{\bar{c}\bar{s}}$ configurations within the framework of QCD sum rules~\cite{Reinders85,Shifman79}. We shall investigate the D-wave tetraquarks in different excitation structures including the $\rho$-mode and $\lambda$-mode. 

This paper is organized as follows. In Sec.~\ref{Sec:Current}, we construct the nonlocal D-wave interpolating currents for $cs\bar{c}\bar{s}$ tetraquark states with $J^{PC}=1^{++}$ and $1^{+-}$ in various excitation structures and color configurations. In Sec.~\ref{Sec:QCDSR}, we introduce the formalism of tetraquark QCD sum rules and calculate the two-point correlation functions and spectral densities for all currents. We perform the numerical analyses to extract the full mass spectra of these D-wave $cs\bar{c}\bar{s}$ tetraquark states in Sec.~\ref{Numerical}. The last section is a summary. 

\section{Interpolating currents for the D-wave $cs\bar{c}\bar{s}$ tetraquarks}\label{Sec:Current}
\begin{figure}[h!]
\centering
\includegraphics[width=5cm]{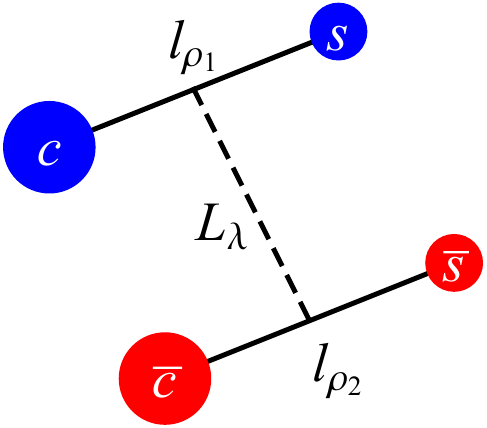}\\
\caption{The excitation structure of the hidden-charm $cs\bar{c}\bar{s}$ tetraquark system, in which $\bf{l}_{\rho_1}$($\bf{l}_{\rho_2}$) is the internal orbital angular momentum for the $cs$($\bar c\bar s$) field, and $\bf{L_\lambda}$ is the orbital angular momentum between the diquark and antidiquark fields.}
\label{FigTetra}
\end{figure}

In this section, we construct the D-wave $cs\bar{c}\bar{s}$ tetraquark interpolating currents with $J^{PC}=1^{++}$ and $1^{+-}$. The $cs\bar{c}\bar{s}$ tetraquark is composed of a $cs$ diquark and a $\bar c\bar s$ antidiquark fields. By analogy with the heavy baryon system, the orbital angular momentum of the tetraquark can be decomposed into $\bf{L}=\bf{L_\rho}+\bf{L_\lambda}=\bf{l}_{\rho_1}+\bf{l}_{\rho_2}+\bf{L_\lambda}$, where $\bf{l}_{\rho_1}$($\bf{l}_{\rho_2}$) is the internal orbital angular momentum for the $cs$($\bar c\bar s$) field, and $\bf{L}_\lambda$ is the orbital angular momentum between the diquark and antidiquark fields. 
It is convenient to denote the orbital excitation of the tetraquark system as $(L_\lambda,L_\rho\{l_{\rho_1},l_{\rho_2}\})$, as shown in Fig.~\ref{FigTetra}. The D-wave excited $cs\bar{c}\bar{s}$ tetraquarks are the excitations with $L_{\rho}+L_\lambda=2$. There exist several different excitation structures for the D-wave tetraquarks: $(L_\lambda,L_\rho\{l_{\rho_1},l_{\rho_2}\})=(2,0\{0,0\})$, $(1,1\{1,0\})$, $(1,1\{0,1\})$, $(0,2\{1,1\})$, $(0,2\{2,0\})$, $(0,2\{0,2\})$. We shall 
study all these D-wave tetraquarks by constructing the interpolating currents with the same structures and quantum numbers. 

\par The color structure of a diquark-antidiquark tetraquark operator $[cs][\bar{c}\bar{s}]$ can be written via $\mathrm{SU(3)}$ symmetry
\begin{equation}\label{colorstructure}
\begin{split}
(\mathbf{3}\otimes\mathbf{3})_{[cs]}\otimes(\bar{\mathbf{3}}\otimes\bar{\mathbf{3}})_{[\bar{c}\bar{s}]}=&(\mathbf{6}\oplus\bar{\mathbf{3}})_{[cs]}\otimes(\mathbf{3}\oplus\bar{\mathbf{6}})_{[\bar{c}\bar{s}]}\\
=&(\mathbf{6}\otimes\bar{\mathbf{6}})\oplus(\bar{\mathbf{3}}\otimes\mathbf{3})\oplus(\mathbf{6}\otimes\mathbf{3})\oplus(\bar{\mathbf{3}}\otimes\bar{\mathbf{6}})\\
=&(\mathbf{1}\oplus\mathbf{8}\oplus\mathbf{27})\oplus(\mathbf{1}\oplus\mathbf{8})\oplus(\mathbf{8}\oplus{\mathbf{10}})\oplus(\mathbf{8}\oplus\overline{\mathbf{10}})\, ,
\end{split}
\end{equation}
in which the color singlet structures come from the $\mathbf{6}_{cs}\otimes\bar{\mathbf{6}}_{\bar{c}\bar{s}}$ and $\bar{\mathbf{3}}_{cs}\otimes\mathbf{3}_{\bar{c}\bar{s}}$ terms, denoting as the color symmetric and antisymmetric configurations respectively. In this work, we consider both these two color configurations. We use only the S-wave good diquark field $\mathcal{O}_S=c^T_{a}\mathbb{C}\gamma_{5}s_{b}$ with $J^P={0^+}$ to compose the D-wave $cs\bar{c}\bar{s}$ tetraquark currents by inserting covariant derivative operators. For example, one can obtain a $\rho$-mode P-wave diquark field with $J^P=1^-$
\begin{align}
\mathcal{O}_{P,\,\mu}=c^T_{a}\mathbb{C}\gamma_{5}D_{\mu}s_{b}\, ,
\end{align}
and a $\rho$-mode D-wave diquark field with $J^P=2^+$
\begin{align}
\mathcal{O}_{D,\,\mu\nu}=c^T_{a}\mathbb{C}\gamma_{5}D_{\mu}D_{\nu}s_{b}\, ,
\end{align}
where $D_{\mu}=\partial_{\mu}+ig_sA_\mu$ is the covariant derivative, the subscripts $a,b$ are color indices, $\mathbb{C}$ denotes the charge conjugate operator, and $T$ represents the transpose of the quark fields. The corresponding charge conjugate antidiquark fields are
\begin{eqnarray}
\nonumber \bar{\mathcal{O}}_S&=&\bar{c}_{a}\mathbb{C}\gamma_5\bar{s}^T_{b},\\
\bar{\mathcal{O}}_{P,\, \mu}&=&\bar{c}_{a}\mathbb{C}\gamma_5 D_\mu \bar{s}^T_{b},\\
\nonumber \bar{\mathcal{O}}_{D,\, \mu\nu}&=&\bar{c}_{a}\mathbb{C}\gamma_5 D_\mu D_\nu\bar{s}^T_{b}.
\end{eqnarray}
To compose the $\lambda$-mode excited tetraquark operator, one should insert the covariant derivative operator between the diquark and antidiquark fields,
\begin{equation}\label{EqLambdaExcitedTetraquark}
\begin{split}
L_\lambda=0&:\qquad \mathcal{O}_S\bar{\mathcal{O}}_S,\\
L_\lambda=1&:\qquad \mathcal{O}_SD_{\mu}\bar{\mathcal{O}}_S,\\
L_\lambda=2&:\qquad \mathcal{O}_SD_\mu D_\nu\bar{\mathcal{O}}_S.\\
\end{split}
\end{equation} 
Considering both the symmetric and antisymmetric color configurations, we construct the D-wave $cs\bar{c}\bar{s}$ interpolating tetraquark currents with 
$J^{PC}=1^{++}$ as 
{\allowdisplaybreaks
\begin{eqnarray}\label{EqCurrent}
\nonumber J_{1,\, \mu\nu}^{A}&=&[c_a^T\mathbb{C}\gamma_5s_b]\{D_\mu,D_\nu\}\left([\bar{c}_a\mathbb{C}\gamma_5\bar{s}_b^T]-[\bar{c}_b\mathbb{C}\gamma_5\bar{s}_a^T]\right)+[\bar{c}_a\mathbb{C}\gamma_5\bar{s}^T_b]\{D_\mu,D_\nu\}\left([c_a^T\mathbb{C}\gamma_5s_b]-[c_b^T\mathbb{C}\gamma_5s_a]\right),\\
\nonumber J_{1,\, \mu\nu}^{S}&=&[c_a^T\mathbb{C}\gamma_5s_b]\{D_\mu,D_\nu\}\left([\bar{c}_a\mathbb{C}\gamma_5\bar{s}_b^T]+[\bar{c}_b\mathbb{C}\gamma_5\bar{s}_a^T]\right)+[\bar{c}_a\mathbb{C}\gamma_5\bar{s}^T_b]\{D_\mu,D_\nu\}\left([c_a^T\mathbb{C}\gamma_5s_b]+[c_b^T\mathbb{C}\gamma_5s_a]\right),\\
\nonumber J_{2,\, \mu\nu}^{A}&=&[c_a^T\mathbb{C}\gamma_5D_\mu s_b]\;D_\nu\left([\bar{c}_a\mathbb{C}\gamma_5\bar{s}_b^T]-[\bar{c}_b\mathbb{C}\gamma_5\bar{s}_a^T]\right)+[\bar{c}_a\mathbb{C}\gamma_5D_\mu\bar{s}^T_b]\;D_\nu\left([c_a^T\mathbb{C}\gamma_5s_b]-[c_b^T\mathbb{C}\gamma_5s_a]\right), \\
\nonumber J_{2,\, \mu\nu}^{S}&=&[c_a^T\mathbb{C}\gamma_5D_\mu s_b]\;D_\nu\left([\bar{c}_a\mathbb{C}\gamma_5\bar{s}_b^T]+[\bar{c}_b\mathbb{C}\gamma_5\bar{s}_a^T]\right)+[\bar{c}_a\mathbb{C}\gamma_5D_\mu\bar{s}^T_b]\;D_\nu\left([c_a^T\mathbb{C}\gamma_5s_b]+[c_b^T\mathbb{C}\gamma_5s_a]\right), \\
\nonumber J_{3,\, \mu\nu}^{A}&=&[c_a^T\mathbb{C}\gamma_5s_b]\;D_\mu\left([\bar{c}_a\mathbb{C}\gamma_5D_\nu\bar{s}_b^T]-[\bar{c}_b\mathbb{C}\gamma_5D_\nu\bar{s}_a^T]\right)+[\bar{c}_a\mathbb{C}\gamma_5\bar{s}^T_b]\;D_\mu\left([c_a^T\mathbb{C}\gamma_5D_\nu s_b]-[c_b^T\mathbb{C}\gamma_5D_\nu s_a]\right), \\
\nonumber J_{3,\, \mu\nu}^{S}&=&[c_a^T\mathbb{C}\gamma_5s_b]\;D_\mu\left([\bar{c}_a\mathbb{C}\gamma_5D_\nu\bar{s}_b^T]+[\bar{c}_b\mathbb{C}\gamma_5D_\nu\bar{s}_a^T]\right)+[\bar{c}_a\mathbb{C}\gamma_5\bar{s}^T_b]\;D_\mu\left([c_a^T\mathbb{C}\gamma_5D_\nu s_b]+[c_b^T\mathbb{C}\gamma_5D_\nu s_a]\right), \\
 J_{4,\, \mu\nu}^{A}&=&[c_a^T\mathbb{C}D_\mu\gamma_5s_b]\left([\bar{c}_a\mathbb{C}\gamma_5D_\nu\bar{s}_b^T]-[\bar{c}_b\mathbb{C}\gamma_5D_\nu\bar{s}_a^T]\right)+[\bar{c}_a\mathbb{C}\gamma_5D_\mu\bar{s}^T_b]\left([c_a^T\mathbb{C}\gamma_5D_\nu s_b]-[c_b^T\mathbb{C}\gamma_5D_\nu s_a]\right), \\
\nonumber J_{4,\, \mu\nu}^{S}&=&[c_a^T\mathbb{C}D_\mu\gamma_5s_b]\left([\bar{c}_a\mathbb{C}\gamma_5D_\nu\bar{s}_b^T]+[\bar{c}_b\mathbb{C}\gamma_5D_\nu\bar{s}_a^T]\right)+[\bar{c}_a\mathbb{C}\gamma_5D_\mu\bar{s}^T_b]\left([c_a^T\mathbb{C}\gamma_5D_\nu s_b]+[c_b^T\mathbb{C}\gamma_5D_\nu s_a]\right), \\
\nonumber J_{5,\, \mu\nu}^{A}&=&[c_a^T\mathbb{C}\gamma_5D_\mu\;D_\nu s_b]\left([\bar{c}_a\mathbb{C}\gamma_5\bar{s}_b^T]-[\bar{c}_b\mathbb{C}\gamma_5\bar{s}_a^T]\right)+[\bar{c}_a\mathbb{C}\gamma_5D_\mu\;D_\nu\bar{s}^T_b]\left([c_a^T\mathbb{C}\gamma_5s_b]-[c_b^T\mathbb{C}\gamma_5s_a]\right),\\
\nonumber J_{5,\, \mu\nu}^{S}&=&[c_a^T\mathbb{C}\gamma_5D_\mu\;D_\nu s_b]\left([\bar{c}_a\mathbb{C}\gamma_5\bar{s}_b^T]+[\bar{c}_b\mathbb{C}\gamma_5\bar{s}_a^T]\right)+[\bar{c}_a\mathbb{C}\gamma_5D_\mu\;D_\nu\bar{s}^T_b]\left([c_a^T\mathbb{C}\gamma_5s_b]+[c_b^T\mathbb{C}\gamma_5s_a]\right),\\
\nonumber J_{6,\, \mu\nu}^{A}&=&[c_a^T\mathbb{C}\gamma_5s_b]\left([\bar{c}_a\mathbb{C}\gamma_5D_\mu\;D_\nu \bar{s}_b^T]-[\bar{c}_b\mathbb{C}\gamma_5D_\mu\;D_\nu \bar{s}_a^T]\right)+[\bar{c}_a\mathbb{C}\gamma_5\bar{s}^T_b]\left([c_a^T\mathbb{C}\gamma_5D_\mu\;D_\nu s_b]-[c_b^T\mathbb{C}\gamma_5D_\mu\;D_\nu s_a]\right), \\
\nonumber J_{6,\, \mu\nu}^{S}&=&[c_a^T\mathbb{C}\gamma_5s_b]\left([\bar{c}_a\mathbb{C}\gamma_5D_\mu\;D_\nu \bar{s}_b^T]+[\bar{c}_b\mathbb{C}\gamma_5D_\mu\;D_\nu \bar{s}_a^T]\right)+[\bar{c}_a\mathbb{C}\gamma_5\bar{s}^T_b]\left([c_a^T\mathbb{C}\gamma_5D_\mu\;D_\nu s_b]+[c_b^T\mathbb{C}\gamma_5D_\mu\;D_\nu s_a]\right),
\end{eqnarray}
}
and the D-wave $cs\bar{c}\bar{s}$ interpolating tetraquark currents with $J^{PC}=1^{+-}$ as 
{\allowdisplaybreaks
\begin{eqnarray}\label{EqCurrent2}
\nonumber J_{7,\, \mu\nu}^{A}&=&[c_a^T\mathbb{C}\gamma_5s_b]\{D_\mu,D_\nu\}\left([\bar{c}_a\mathbb{C}\gamma_5\bar{s}_b^T]-[\bar{c}_b\mathbb{C}\gamma_5\bar{s}_a^T]\right)-[\bar{c}_a\mathbb{C}\gamma_5\bar{s}^T_b]\{D_\mu,D_\nu\}\left([c_a^T\mathbb{C}\gamma_5s_b]-[c_b^T\mathbb{C}\gamma_5s_a]\right),\\
\nonumber J_{7,\, \mu\nu}^{S}&=&[c_a^T\mathbb{C}\gamma_5s_b]\{D_\mu,D_\nu\}\left([\bar{c}_a\mathbb{C}\gamma_5\bar{s}_b^T]+[\bar{c}_b\mathbb{C}\gamma_5\bar{s}_a^T]\right)-[\bar{c}_a\mathbb{C}\gamma_5\bar{s}^T_b]\{D_\mu,D_\nu\}\left([c_a^T\mathbb{C}\gamma_5s_b]+[c_b^T\mathbb{C}\gamma_5s_a]\right),\\
\nonumber J_{8,\, \mu\nu}^{A}&=&[c_a^T\mathbb{C}\gamma_5D_\mu s_b]\;D_\nu\left([\bar{c}_a\mathbb{C}\gamma_5\bar{s}_b^T]-[\bar{c}_b\mathbb{C}\gamma_5\bar{s}_a^T]\right)-[\bar{c}_a\mathbb{C}\gamma_5D_\mu\bar{s}^T_b]\;D_\nu\left([c_a^T\mathbb{C}\gamma_5s_b]-[c_b^T\mathbb{C}\gamma_5s_a]\right), \\
\nonumber J_{8,\, \mu\nu}^{S}&=&[c_a^T\mathbb{C}\gamma_5D_\mu s_b]\;D_\nu\left([\bar{c}_a\mathbb{C}\gamma_5\bar{s}_b^T]+[\bar{c}_b\mathbb{C}\gamma_5\bar{s}_a^T]\right)-[\bar{c}_a\mathbb{C}\gamma_5D_\mu\bar{s}^T_b]\;D_\nu\left([c_a^T\mathbb{C}\gamma_5s_b]+[c_b^T\mathbb{C}\gamma_5s_a]\right), \\
\nonumber J_{9,\, \mu\nu}^{A}&=&[c_a^T\mathbb{C}\gamma_5s_b]\;D_\mu\left([\bar{c}_a\mathbb{C}\gamma_5D_\nu\bar{s}_b^T]-[\bar{c}_b\mathbb{C}\gamma_5D_\nu\bar{s}_a^T]\right)-[\bar{c}_a\mathbb{C}\gamma_5\bar{s}^T_b]\;D_\mu\left([c_a^T\mathbb{C}\gamma_5D_\nu s_b]-[c_b^T\mathbb{C}\gamma_5D_\nu s_a]\right), \\
\nonumber J_{9,\, \mu\nu}^{S}&=&[c_a^T\mathbb{C}\gamma_5s_b]\;D_\mu\left([\bar{c}_a\mathbb{C}\gamma_5D_\nu\bar{s}_b^T]+[\bar{c}_b\mathbb{C}\gamma_5D_\nu\bar{s}_a^T]\right)-[\bar{c}_a\mathbb{C}\gamma_5\bar{s}^T_b]\;D_\mu\left([c_a^T\mathbb{C}\gamma_5D_\nu s_b]+[c_b^T\mathbb{C}\gamma_5D_\nu s_a]\right), \\
J_{10,\, \mu\nu}^{A}&=&[c_a^T\mathbb{C}D_\mu\gamma_5s_b]\left([\bar{c}_a\mathbb{C}\gamma_5D_\nu\bar{s}_b^T]-[\bar{c}_b\mathbb{C}\gamma_5D_\nu\bar{s}_a^T]\right)-[\bar{c}_a\mathbb{C}\gamma_5D_\mu\bar{s}^T_b]\left([c_a^T\mathbb{C}\gamma_5D_\nu s_b]-[c_b^T\mathbb{C}\gamma_5D_\nu s_a]\right), \\
\nonumber J_{10,\, \mu\nu}^{S}&=&[c_a^T\mathbb{C}D_\mu\gamma_5s_b]\left([\bar{c}_a\mathbb{C}\gamma_5D_\nu\bar{s}_b^T]+[\bar{c}_b\mathbb{C}\gamma_5D_\nu\bar{s}_a^T]\right)-[\bar{c}_a\mathbb{C}\gamma_5D_\mu\bar{s}^T_b]\left([c_a^T\mathbb{C}\gamma_5D_\nu s_b]+[c_b^T\mathbb{C}\gamma_5D_\nu s_a]\right), \\
\nonumber J_{11,\, \mu\nu}^{A}&=&[c_a^T\mathbb{C}\gamma_5D_\mu\;D_\nu s_b]\left([\bar{c}_a\mathbb{C}\gamma_5\bar{s}_b^T]-[\bar{c}_b\mathbb{C}\gamma_5\bar{s}_a^T]\right)-[\bar{c}_a\mathbb{C}\gamma_5D_\mu\;D_\nu\bar{s}^T_b]\left([c_a^T\mathbb{C}\gamma_5s_b]-[c_b^T\mathbb{C}\gamma_5s_a]\right),\\
\nonumber J_{11,\, \mu\nu}^{S}&=&[c_a^T\mathbb{C}\gamma_5D_\mu\;D_\nu s_b]\left([\bar{c}_a\mathbb{C}\gamma_5\bar{s}_b^T]+[\bar{c}_b\mathbb{C}\gamma_5\bar{s}_a^T]\right)-[\bar{c}_a\mathbb{C}\gamma_5D_\mu\;D_\nu\bar{s}^T_b]\left([c_a^T\mathbb{C}\gamma_5s_b]+[c_b^T\mathbb{C}\gamma_5s_a]\right),\\
\nonumber J_{12,\, \mu\nu}^{A}&=&[c_a^T\mathbb{C}\gamma_5s_b]\left([\bar{c}_a\mathbb{C}\gamma_5D_\mu\;D_\nu \bar{s}_b^T]-[\bar{c}_b\mathbb{C}\gamma_5D_\mu\;D_\nu \bar{s}_a^T]\right)-[\bar{c}_a\mathbb{C}\gamma_5\bar{s}^T_b]\left([c_a^T\mathbb{C}\gamma_5D_\mu\;D_\nu s_b]-[c_b^T\mathbb{C}\gamma_5D_\mu\;D_\nu s_a]\right), \\
\nonumber J_{12,\, \mu\nu}^{S}&=&[c_a^T\mathbb{C}\gamma_5s_b]\left([\bar{c}_a\mathbb{C}\gamma_5D_\mu\;D_\nu \bar{s}_b^T]+[\bar{c}_b\mathbb{C}\gamma_5D_\mu\;D_\nu \bar{s}_a^T]\right)-[\bar{c}_a\mathbb{C}\gamma_5\bar{s}^T_b]\left([c_a^T\mathbb{C}\gamma_5D_\mu\;D_\nu s_b]+[c_b^T\mathbb{C}\gamma_5D_\mu\;D_\nu s_a]\right),
\end{eqnarray}
}
where $\{D_\mu,D_\nu\}=D_\mu D_\nu+D_\nu D_\mu$. The interpolating currents with the superscripts ``$S$" and ``$A$" denote the symmetric $[cs]_{\mathbf{6}}[\bar{c}\bar{s}]_{\bar{\mathbf{6}}}$ and antisymmetric $[cs]_{\bar{\mathbf{3}}}[\bar{c}\bar{s}]_{\mathbf{3}}$ color structures, which are abbreviated to $\mathbf{3}$ and $\mathbf{6}$ respectively in the following. We show the excitation structures $(L_\lambda,L_\rho\{l_{\rho_1},l_{\rho_2}\})$, color configurations and $J^{PC}$ quantum numbers for these interpolating currents in Tabel~\ref{TabCurrent}. The abbreviation $\mathbf{3}_{\lambda\lambda}/\mathbf{6}_{\lambda\lambda}$ ($\mathbf{3}_{\rho\rho}/\mathbf{6}_{\rho\rho}$) indicates that the corresponding current contains two $\lambda$-orbital ($\rho$-orbital) momentums with antisymmetric/symmetric color structure, while $\mathbf{3}_{\lambda\rho}/\mathbf{6}_{\lambda\rho}$ indicates that the current contains one $\lambda$-orbital momentum and one $\rho$-orbital momentum with antisymmetric/symmetric color structure. In the following, we shall investigate the mass spectra for the D-wave $cs\bar{c}\bar{s}$ tetraquarks by using these interpolating currents. For the currents belong to the $(0,2\{2,0\})$ and $(0,2\{0,2\})$ structures, we only study the $(0,2\{2,0\})$ ones because the $(0,2\{0,2\})$ currents will provide the same results in our calculations. 
\begin{table}[h!]
\caption{The excitation structures, color configurations and $J^{PC}$ quantum numbers for the D-wave $cs\bar{c}\bar{s}$ interpolating currents in Eqs.~\eqref{EqCurrent}-\eqref{EqCurrent2}.}\label{TabCurrent}
\renewcommand\arraystretch{1.5} 
\setlength{\tabcolsep}{2.em}{ 
\begin{tabular}{c c c c}
  \hline
   \hline
$(L_\lambda,L_\rho\{l_{\rho_1},l_{\rho_2}\})$ & $[cs]_{\bar{\mathbf{3}}}[\bar{c}\bar{s}]_{\mathbf{3}}$ & $[cs]_{\mathbf{6}}[\bar{c}\bar{s}]_{\bar{\mathbf{6}}}$ & $J^{PC}$ \\
  \hline
$(2,0\{0,0\})$ &  $J_{1,\, \mu\nu}^{A}(\mathbf{3}_{\lambda\lambda})$ & $J_{1,\, \mu\nu}^{S}(\mathbf{6}_{\lambda\lambda})$ & $1^{++}$ \vspace{1ex}\\
    &  $J_{7,\, \mu\nu}^{A}(\mathbf{3}_{\lambda\lambda})$ & $J_{7,\, \mu\nu}^{S}(\mathbf{6}_{\lambda\lambda})$ & $1^{+-}$ \vspace{1ex}\\
$(1,1\{1,0\})$ &  $J_{2,\, \mu\nu}^{A}(\mathbf{3}_{\lambda\rho})$ & $J_{2,\, \mu\nu}^{S}(\mathbf{6}_{\lambda\rho})$ & $1^{++}$ \vspace{1ex}\\
                      &  $J_{8,\, \mu\nu}^{A}(\mathbf{3}_{\lambda\rho})$ & $J_{8,\, \mu\nu}^{S}(\mathbf{6}_{\lambda\rho})$ & $1^{+-}$ \vspace{1ex}\\
$(1,1\{0,1\})$ &  $J_{3,\, \mu\nu}^{A}(\mathbf{3}_{\lambda\rho})$ & $J_{3,\, \mu\nu}^{S}(\mathbf{6}_{\lambda\rho})$ & $1^{++}$ \vspace{1ex}\\
                      &  $J_{9,\, \mu\nu}^{A}(\mathbf{3}_{\lambda\rho})$ & $J_{9,\, \mu\nu}^{S}(\mathbf{6}_{\lambda\rho})$ & $1^{+-}$ \vspace{1ex}\\
$(0,2\{1,1\})$ &  $J_{4,\, \mu\nu}^{A}(\mathbf{3}_{\rho\rho})$ & $J_{4,\, \mu\nu}^{S}(\mathbf{6}_{\rho\rho})$ & $1^{++}$ \vspace{1ex}\\
                      &  $J_{10,\, \mu\nu}^{A}(\mathbf{3}_{\rho\rho})$ & $J_{10,\, \mu\nu}^{S}(\mathbf{6}_{\rho\rho})$ & $1^{+-}$ \vspace{1ex}\\
$(0,2\{2,0\})$ &  $J_{5,\, \mu\nu}^{A}(\mathbf{3}_{\rho\rho})$ & $J_{5,\, \mu\nu}^{S}(\mathbf{6}_{\rho\rho})$ & $1^{++}$ \vspace{1ex}\\
                      &  $J_{11,\, \mu\nu}^{A}(\mathbf{3}_{\rho\rho})$ & $J_{11,\, \mu\nu}^{S}(\mathbf{6}_{\rho\rho})$ & $1^{+-}$ \vspace{1ex}\\
$(0,2\{0,2\})$ &  $J_{6,\, \mu\nu}^{A}(\mathbf{3}_{\rho\rho})$ & $J_{6,\, \mu\nu}^{S}(\mathbf{6}_{\rho\rho})$ & $1^{++}$ \vspace{1ex}\\
                      &  $J_{12,\, \mu\nu}^{A}(\mathbf{3}_{\rho\rho})$ & $J_{12,\, \mu\nu}^{S}(\mathbf{6}_{\rho\rho})$ & $1^{+-}$ \vspace{1ex}\\                                                                                        
  \hline
   \hline
\end{tabular}
}
\end{table}

\section{QCD Sum Rules for tetraquark states}\label{Sec:QCDSR}
In this section, we shall introduce the method of QCD sum rules for the hidden-charm tetraquark states. The two-point correlation functions for the tensor currents can be written as
\begin{equation}\label{correlation1}
\begin{split}
\Pi_{\mu\nu,\,\rho\sigma}(q^2)&=i\int d^4x e^{iq\cdot x}\langle 0|T\left[J_{\mu\nu}(x)J^{\dagger}_{\rho\sigma}(0)\right]|0\rangle\\
                            &=T^+_{\mu\nu\rho\sigma}\Pi_1(q^2)+\cdots \, ,
\end{split}
\end{equation}
where
\begin{equation}
\begin{split}
T^\pm_{\mu\nu\rho\sigma}&=\left(\frac{q_\mu q_\nu}{q^2}\eta_{\nu\sigma}\pm(\mu\leftrightarrow\nu)\right)\pm(\rho\leftrightarrow\sigma),\\
\eta_{\mu\nu}&=\frac{q_{\mu} q_{\nu}}{q^{2}}-g_{\mu \nu},
\end{split}
\end{equation} 
and $\Pi_{1}\left(q^{2}\right)$ is the polarization function related to the spin-1 intermediate state, the $``\cdots"$ represents other tensor structures relating to different hadron states.
The tensor current can couple to the spin-1 physical state $X$ through
\begin{equation}\label{coupling}
\begin{split}
\langle0|J_{\mu\nu}(x)|1^{\mathbb{P}\mathbb{C}}(\overrightarrow{\mathbf{p}},r)\rangle&=Z\epsilon^{\mu\nu\alpha\beta}\in_\alpha(\overrightarrow{\mathbf{p}},r)p_\beta,\\
\langle0|J_{\mu\nu}(x)|1^{(-\mathbb{P})\mathbb{C}}(\overrightarrow{\mathbf{p}},r)\rangle&=Z_+(\in^\mu(\overrightarrow{\mathbf{p}},r)p^\nu+\in^\nu(\overrightarrow{\mathbf{p}},r)p^\mu)\\
&\quad+Z_-(\in^\mu(\overrightarrow{\mathbf{p}},r)p^\nu-\in^\nu(\overrightarrow{\mathbf{p}},r)p^\mu),
\end{split}
\end{equation}
where $Z, Z_+, Z_-$ are coupling constants, $\epsilon^{\mu\nu\alpha\beta}$ is the antisymmetical tensor and $\in_\alpha$ is the polarization tensor.

At the hadron level, the two-point correlation function can be written as
\begin{equation}
\Pi(q^2)=\frac{1}{\pi}\int^{\infty}_{s_<}\frac{\mathrm{Im}\Pi(s)}{s-q^2-i\epsilon}ds,
\end{equation}
where we have used the form of the dispersion relation, and $s_<$ denotes the physical threshold. The imaginary part of the correlation function is defined as the spectral function, which is usually evaluated at the hadron level by inserting intermediate hadron states $\sum_n|n\rangle\langle n|$
\begin{equation}\label{spectral}
\begin{split}
\rho(s)\equiv\frac{1}{\pi}\mathrm{Im}\Pi(s)&=\sum_n\delta(s-M^2_n)\langle 0|\eta|n\rangle\langle n|\eta^\dag|0\rangle\\
&=f^2_X\delta(s-m^2_X)+\mathrm{continuum},
\end{split}
\end{equation}
where we have adopted the usual parametrization of one-pole dominance for the ground state $X$ and a continuum contribution. One can find some investigations of the excited mesons~\cite{Reinders:1981si,Aliev:1981ju,Dai:1996yw}, baryons~\cite{Chen:2015kpa} and tetraquarks~\cite{Zhang:2010mv,Zhang:2010mw,Chen:2016oma} in QCD sum rules by using the non-local interpolating currents under the ``pole+continuum'' approximation. The spectral density $\rho(s)$ can also be evaluated at the quark-gluon level via the operator product expansion(OPE). 
To pick out the contribution of the lowest lying resonance in \eqref{spectral}, the QCD sum rules are established as 
\begin{equation}\label{EqL}
\mathcal{L}_k(s_0,M_B^2)=f^2_Xm^{2k}_{H}e^{-m^2_H/M_B^2}=\int^{s_0}_{4m_c^2}ds\,e^{-s/M_B^2}\rho(s)s^k
\end{equation}
in which $M_B$ represents the Borel mass introduced by the Borel transformation and $s_0$ is the continuum threshold. The mass of the lowest-lying hadron can be thus extracted as
\begin{equation}\label{ratio}
m_X(s_0,M_B^2)=\sqrt{\frac{\mathcal{L}_1(s_0,M_B^2)}{\mathcal{L}_0(s_0,M_B^2)}},
\end{equation}
which is the function of two parameters $M_B^2$ and $s_0$. We shall discuss the detail to obtain suitable parameter working regions in QCD sum rule analyses in next section.
Using the operator production expansion method, the two-point function can also be evaluated at the quark-gluonic level as a function of various QCD parameters, such as QCD condensates, quark masses and the strong coupling constant $\alpha_s$. To evaluate the Wilson coefficients, we adopt the heavy quark propagator in momentum space and the strange quark propagator in coordinate space 
\begin{eqnarray}
\nonumber i S_{c}^{a b}(p)&=&\frac{i \delta^{a b}}{\hat{p}-m_{c}}
 +\frac{i}{4} g_{s} \frac{\lambda_{a b}^{n}}{2} G_{\mu \nu}^{n} \frac{\sigma^{\mu \nu}\left(\hat{p}+m_{c}\right)+\left(\hat{p}+m_{c}\right) \sigma^{\mu \nu}}{12}
 \\
 & &+\frac{i \delta^{a b}}{12}\left\langle g_{s}^{2} G G\right\rangle m_{c} \frac{p^{2}+m_{c} \hat{p}}{(p^{2}-m_{c}^{2})^{4}}\, , \\
\nonumber i S_{s}^{ab}(x)&=&\frac{i\delta^{ab}}{2\pi^2x^4}\hat{x}-\frac{\delta^{ab}}{12}\langle\bar{s}s\rangle+\frac{i}{32\pi^2}\frac{\lambda^n_{ab}}{2}g_sG^n_{\mu\nu}\frac{1}{x^2}(\sigma^{\mu\nu}\hat{x}+\hat{x}\sigma^{\mu\nu})\\
\nonumber & &+\frac{\delta^{ab}x^2}{192}\langle\bar{s}g_s\sigma\cdot Gs\rangle-\frac{m_s\delta^{ab}}{4\pi^2x^2}+\frac{i\delta^{ab}m_s\langle\bar{s}s\rangle}{48}\hat{x}-\frac{im_s\langle\bar{s}g_s\sigma\cdot Gs\rangle\delta^{ab}x^2\hat{x}}{1152}\, ,
\end{eqnarray}
where $\hat{p}=p^{\mu}\gamma_{\mu}$, $\hat{x}=x^{\mu}\gamma_{\mu}$. In this work, we will evaluate Wilson coefficients of the correlation function up to dimension ten condensates at the leading order of $\alpha_s$. We find that the calculations are really complicate due to the existence of the covariant derivative operators. The results of spectral functions are too lengthy to show here, and thus we list all of them in the Appendix.

\section{Mass sum rule Analyses}\label{Numerical}
In this section we perform the QCD sum rule analyses for the $cs\bar{c}\bar{s}$ tetraquark systems. We use the following values of the quark masses and various QCD condensates~\cite{ParticleDataGroup:2022pth,Narison:1989aq,Jamin:2001zr,Jamin:1998ra,Ioffe:1981kw,Chung:1984gr,Dosch:1988vv,Khodjamirian:2011ub,ParticleDataGroup:2018ovx,Francis:2018jyb}:
\begin{equation}\label{inputparameter}
\begin{split}
  &m_c(m_c)=1.27\pm0.02\;\mathrm{GeV},\\
  &m_c/m_s=11.76^{+0.05}_{-0.10} \, , \\
  &\langle \bar{q}q\rangle=-(0.24\pm0.03)^3\;\mathrm{GeV}^3,\\
  &\langle \bar{q}g_s\sigma\cdot Gq\rangle=-M_0^2\langle \bar{q}q\rangle,\\
  &\langle \bar{q}q\bar{q}q\rangle=\langle \bar{q}q\rangle^2\,,\\
  &M_0^2=(0.8\pm0.2)\;\mathrm{GeV}^2,\\
  &\langle \bar{s}s\rangle/\langle \bar{q}q\rangle=0.8\pm0.1,\\
  &\langle g_s^2GG\rangle=(0.48\pm0.14)\;\mathrm{GeV}^4,\\ 
\end{split}
\end{equation}
where the charm quark mass $m_c$ is the ``running'' mass in the $\overline{\text{MS}}$ scheme. To ensure the unified renormalization scale in our analyses, we use the renormalization scheme and scale independent $m_c/m_s$ mass ratio from PDG~\cite{ParticleDataGroup:2022pth} to obtain the strange quark mass $m_s$.

To establish a stable mass sum rule, one should find the appropriate parameter working regions at first, i.e, for the continuum threshold $s_0$ and the Borel mass $M_B^2$. The threshold $s_0$ can be determined via the minimized variation of the hadronic mass $m_X$ with respect to the Borel mass $M_B^2$. The lower bound on the Borel mass $M_B^2$ can be fixed by requiring a reasonable OPE convergence, while its upper bound is determined through a sufficient pole contribution. The pole contribution is defined as
\begin{equation}\label{EqPC}
\mathrm{PC}(s_0,M_B^2)=\frac{\mathcal{L}_0(s_0,M_B^2)}{\mathcal{L}_0(\infty,M_B^2)},
\end{equation}
where $\mathcal{L}_0$ has been defined in Eq.~\eqref{EqL}.
\begin{figure}[t!]
\centering
\includegraphics[width=10cm]{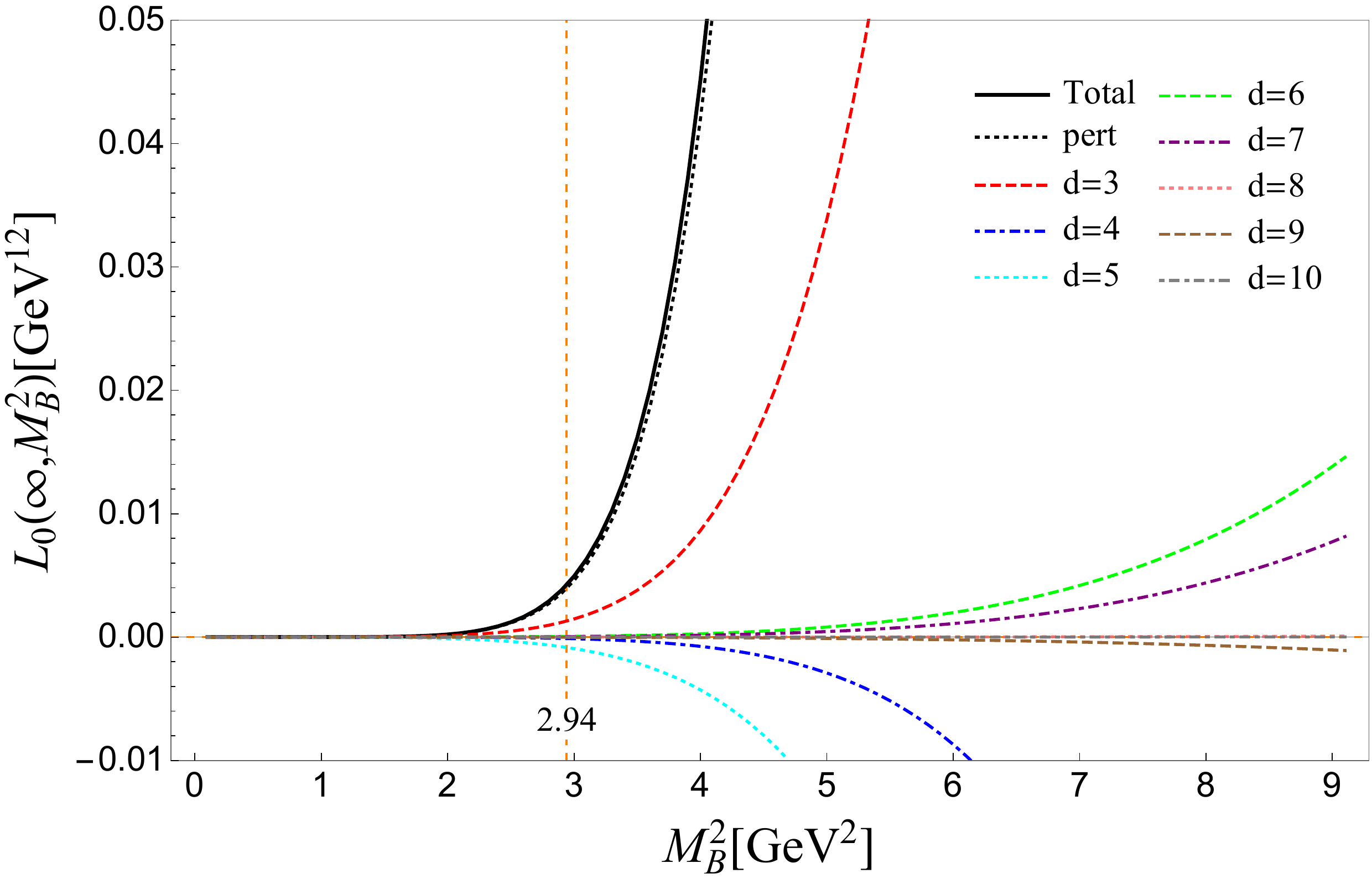}\\
\caption{Contributions of various OPE terms to the correlation function for the current $J_{5,\mu\nu}^{A}(x)$, as a function of $M_B^2$ when $s_0\to\infty$.}
\label{FigRho0220}
\end{figure}


As an example, we use the color antisymmetric current $J_{5,\mu\nu}^{A}(x)$ with $J^{PC}=1^{++}$ in the $(0,2\{2,0\})$ excitation mode to show the details of the numerical analysis. For this current, the dominant non-perturbative contribution to the correlation function comes from the quark condensate, which is proportional to the charm quark mass $m_c$. In Fig.~\ref{FigRho0220}, we show the contributions of the perturbative term and various condensate terms to the correlation function with respect to $M_B^2$ when $s_0$ tends to infinity. It is clear that the Borel mass $M_B^2$ should be large enough to ensure the convergence of the OPE series. In this work, we require that the perturbative term be two times larger than the quark condensate term, providing the lower bound of the Borel mass $M_B^2\geq2.82\;\mathrm{GeV}^2$. The other QCD condensates are much smaller than the quark condensate in this region of $M_B^2$. After studying the pole contribution defined in Eq.~\eqref{EqPC}, one finds that the PC is very small for such D-wave $cs\bar{c}\bar{s}$ tetraquark system due to the high dimension of the interpolating current. To find an upper bound on the Borel mass,  we require that the pole contribution to be larger than $20\%$. As a result, the reasonable Borel window for the current $J_{5,\mu\nu}^{A}(x)$ is obtained as $2.94\;\mathrm{GeV}^2\leq M_B^2\leq3.90\;\mathrm{GeV}^2$. 
\begin{figure}[t!]
\centering
\includegraphics[width=8cm]{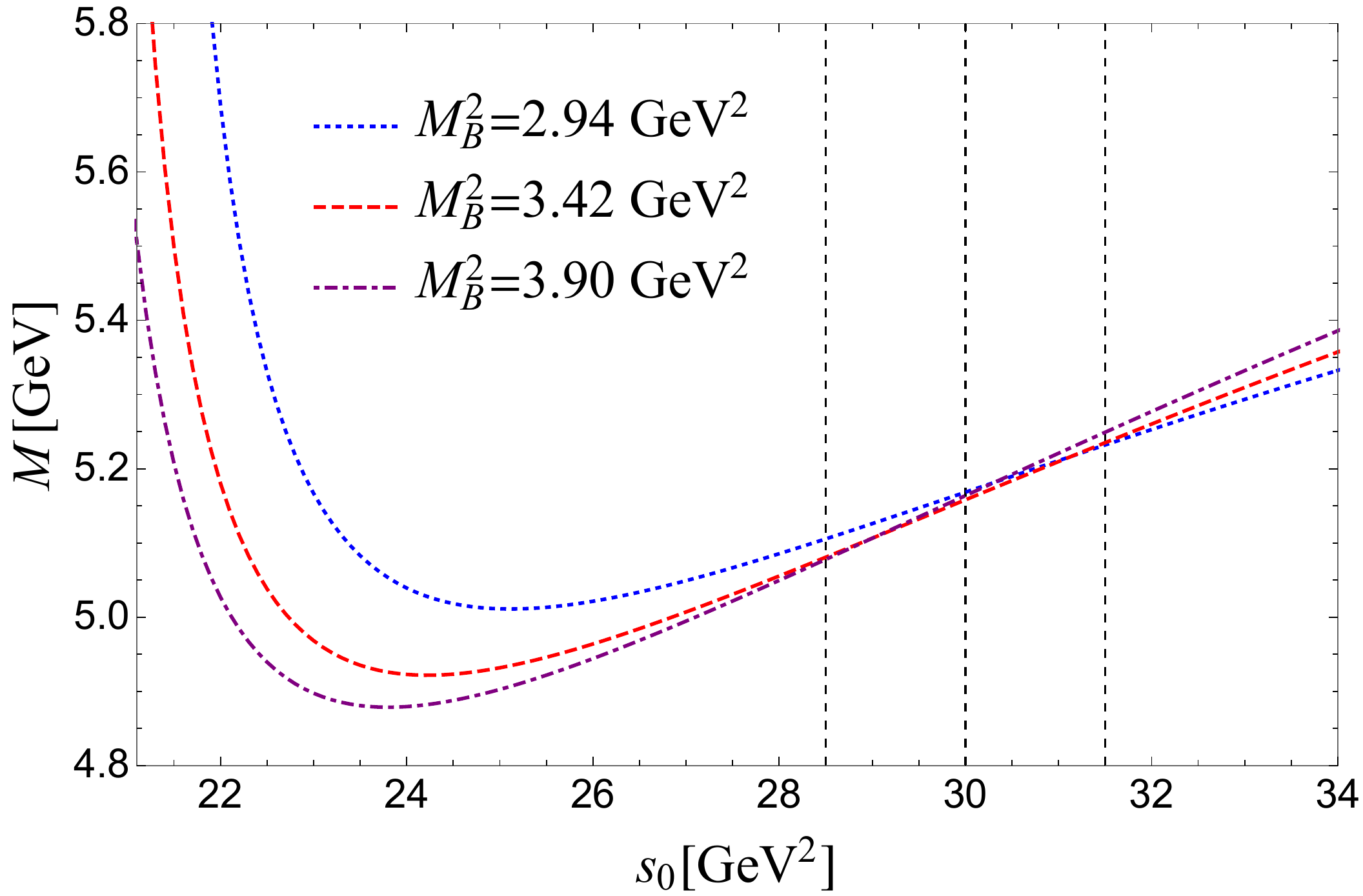}\quad
\includegraphics[width=8cm]{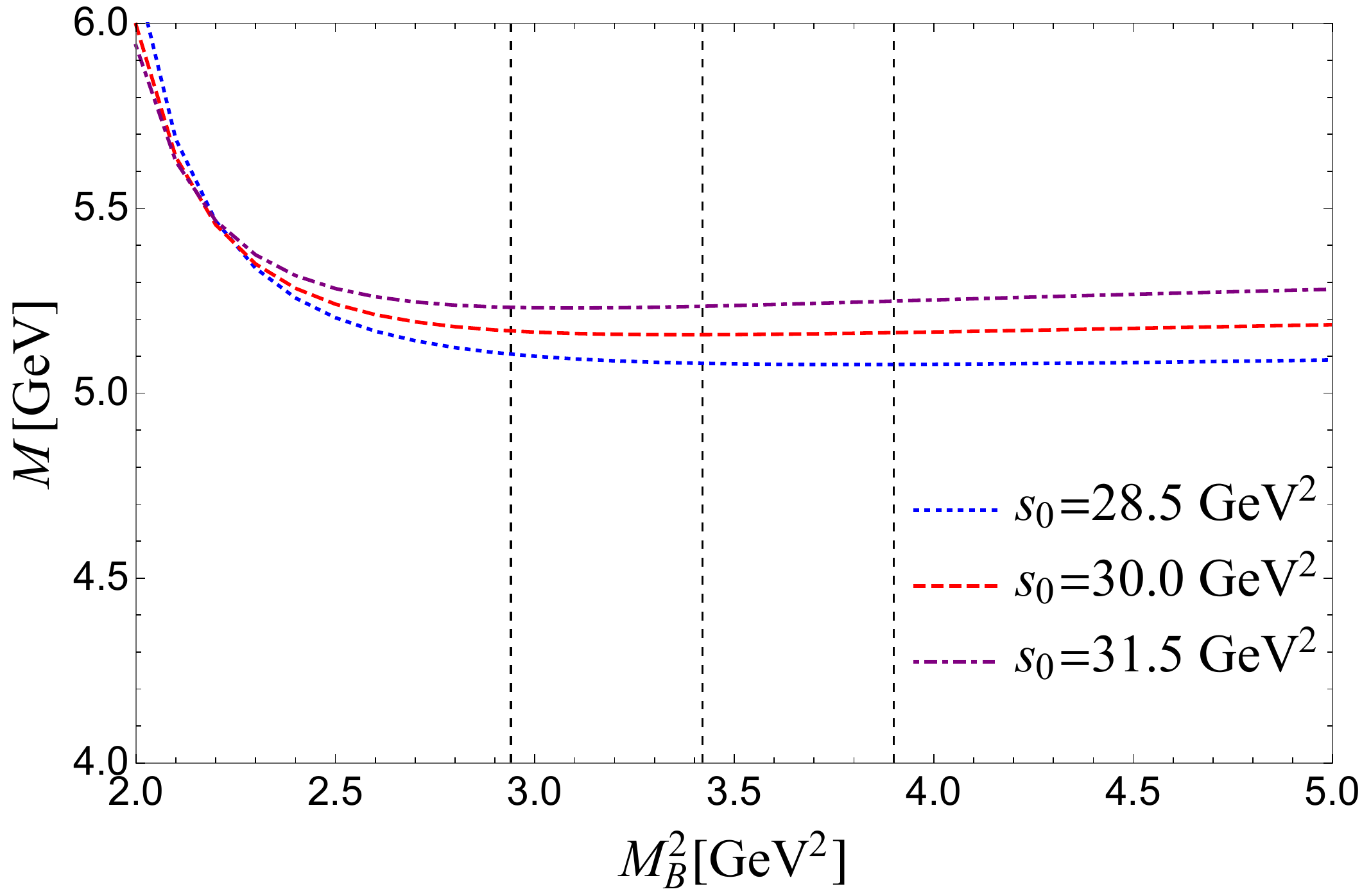}\\
\caption{Mass curves for the interpolating current $J_{5,\mu\nu}^A(x)$ with $J^{PC}=1^{++}$.}
\label{FigmX}
\end{figure}

As mentioned above, the variation of the extracted hadron mass $m_X$ with respect to $M_B^2$ should be minimized to obtain the optimal value of the continuum threshold $s_0$. We show the variation of $m_X$ with $s_0$ in the left panel of Fig.~\ref{FigmX}, from which the optimized value of the continuum threshold can be chosen as $s_0\approx(30.0\pm1.5)\;\mathrm{GeV}^2$. In the right panel of Fig.~\ref{FigmX}, the mass sum rules are established to be very stable in the above parameter regions of $s_0$ and $M_B^2$. The hadron mass for this D-wave $cs\bar{c}\bar{s}$ tetraquark with $J^{PC}=1^{++}$ can be obtained as 
\begin{equation}
m_{J_{5}^A}=5.16_{-0.13}^{+0.12} ~\text{GeV}\,, 
\end{equation}
where the errors come from the uncertainties of the threshold $s_0$, Borel mass $M_B^2$, quark masses and various QCD condensates in Eq.~\eqref{inputparameter}. Performing the same numerical analyses to all interpolating currents in Eqs.~\eqref{EqCurrent}-\eqref{EqCurrent2}, we find that only the current $J_{5,\mu\nu}^{S}(x)$, $J_{11,\mu\nu}^{A(S)}(x)$ and $J_{4,\mu\nu}^{A(S)}(x)$ with $J^{PC}=1^{++}$ shows the same mass sum rule behaviors as $J_{5,\mu\nu}^A(x)$. We collect their numerical results in Table~\ref{cscsResultTab}.

\begin{figure}[t!]
\centering
\includegraphics[width=8cm]{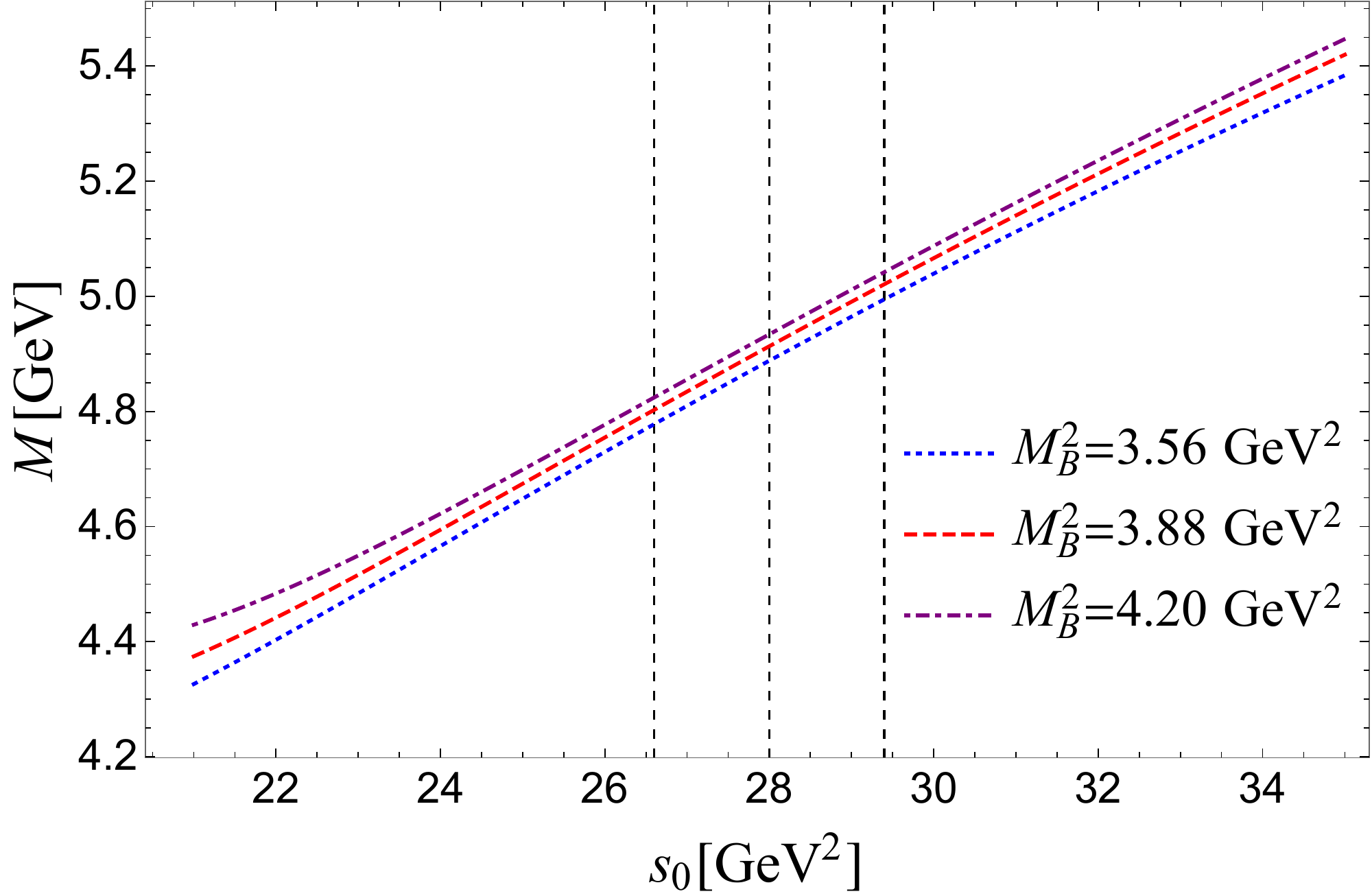}\quad
\includegraphics[width=8cm]{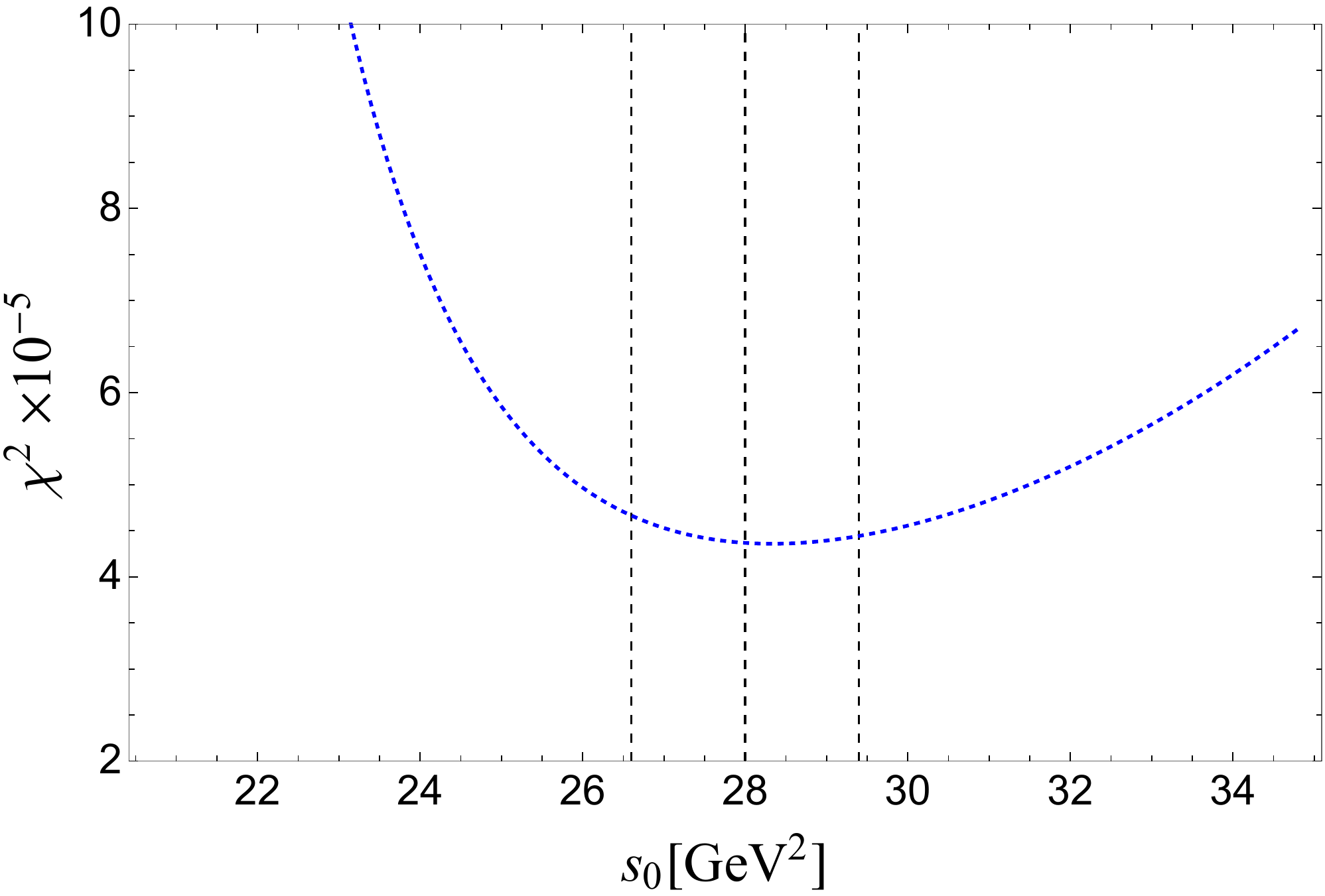}\\
\caption{Mass curves(left) and $\chi^2$ curve(right) for the current $J_{1,\mu\nu}^S(x)$ with $J^{PC}=1^{++}$.}
\label{FigmX2000}
\end{figure}

Except for $J_{5,\mu\nu}^{A(S)}(x)$, $J_{11,\mu\nu}^{A(S)}(x)$ and $J_{4,\mu\nu}^{A(S)}(x)$, the other interpolating currents display very different mass sum rule behavior. As shown in the left panel for $J_{1,\mu\nu}^S(x)$, the extracted hadron mass increases monotonically with the continuum threshold $s_0$. Thus one is not able to find an optimal value of $s_0$ to minimize the variation of hadron mass with respect to $M_B^2$. For such a situation, we define the following hadron mass $\bar{m}_X$ and quantity $\chi^2(s_0)$ to study the stability of mass sum rules
\begin{eqnarray}
\bar{m}_X(s_0)&=&\sum^N_{i=1}\frac{m_X(s_0,M_{B,i}^2)}{N},\\
\chi^2(s_0)&=&\sum^N_{i=1}\left[\frac{m_X(s_0,M_{B,i}^2)}{\bar{m}_X(s_0)}-1\right]^2,
\end{eqnarray}
where the $M_{B,i}^2(i=1,2,\dots,N)$ represent $N$ definite values for the Borel parameter $M_B^2$ in the Borel window. According to the above definition, the optimal choice for the continuum threshold $s_0$ in the QCD sum rule analysis can be obtained by minimizing the quantity $\chi^2(s_0)$, which is only the function of $s_0$. We show this relation in the right panel of Fig.~\ref{FigmX2000}, in which there is a minimum point around $s_0\approx28.0\;\mathrm{GeV}^2$. We can thus determine the working range for the continuum threshold to be $s_0=(28.0\pm1.4)\;\mathrm{GeV}^2$, as shown in the left panel of Fig.~\ref{FigmX2000}. The hadron mass is thus obtained as 
\begin{equation}
m_{J_{1}^S}=4.91_{-0.12}^{+0.11} ~\text{GeV}.
\end{equation}
In these analyses, we find that the OPE series for the $J_{4,\mu\nu}^{A(S)}(x)$ and $J_{10,\mu\nu}^{A(S)}(x)$ belonging to the $(0,2\{1,1\})$ structure are much different from those of other interpolating currents. As shown in the Appendix, the quark condensate gives no contribution to the correlation function for all the $(0,2\{1,1\})$ currents.

After performing similar analyses, we obtain the numerical results for all the other interpolating currents in Eqs.~\eqref{EqCurrent}-\eqref{EqCurrent2} and collect them in Table~\ref{cscsResultTab}. One finds that the extracted hadron masses from $J_{1,\mu\nu}^{A}(x)$ and $J_{5,\mu\nu}^{S}(x)$ with $J^{PC}=1^{++}$ are in good agreement with the mass of the newly observed resonance $X(4685)$, implying that the $X(4685)$ could be interpreted as a D-wave $cs\bar{c}\bar{s}$ tetraquark state with $J^{PC}=1^{++}$ in the excitation mode of $(2,0\{0,0\})$ or $(0,2\{2,0\})$.

Considering the same physical picture for the $(1,1\{1,0\})$ and $(1,1\{0,1\})$ excitation structures, the interpolating currents $J_{2,\mu\nu}^{A(S)}(x)$ and $J_{3,\mu\nu}^{A(S)}(x)$ display similar mass sum rules. The currents $J_{2,\mu\nu}^{A}(x)$ and $J_{3,\mu\nu}^{A}(x)$ give almost degenerate hadron masses, as shown in Table~\ref{cscsResultTab}. To study their mixing effects, we also perform analyses for the mixed currents $J_{2,\mu\nu}^{A(S)}+J_{3,\mu\nu}^{A(S)}$. Our calculations show that the off-diagonal correlator $\Pi_{23}^{A(S)}(q^2)$ is nonzero, implying that the currents $J_{2,\mu\nu}^{A}(x)$ and $J_{3,\mu\nu}^{A}(x)$ may couple to the same hadron state. The same situation also happens for the interpolating currents $J_{8,\mu\nu}^{A(S)}(x)$ and $J_{9,\mu\nu}^{A(S)}(x)$ that they couple to the same tetraquark state. 

\begin{table}[h!]
\caption{The hadron masses of the $cs\bar{c}\bar{s}$ tetraquark states with different $J^{PC}$ quantum numbers and $(L_\lambda,L_\rho\{l_{\rho_1},l_{\rho_2}\})$ excitation structures. The subscripts ``A'' and ``S'' denote the numerical results for the color antisymmetric and symmetric currents, respectively.}\label{cscsResultTab}
\renewcommand\arraystretch{1.6} 
\setlength{\tabcolsep}{0.4 em}{ 
\begin{tabular}{c c c| c c c c |c c c c}
  \hline
  \hline
 $(L_\lambda,L_\rho\{l_{\rho_1},l_{\rho_2}\})$ &\text{Current}& $J^{PC}$& $m_{A}[\mathrm{GeV}]$ & $s_{0,A}[\mathrm{GeV}^2]$ & $M_{B,A}^2[\mathrm{GeV}^2]$ &$PC_{A}[\%]$& $m_{S}[\mathrm{GeV}]$ & $s_{0,S}[\mathrm{GeV}^2]$ & $M_{B,S}^2[\mathrm{GeV}^2]$ &$PC_{S}[\%]$\\
  \hline

  $(2,0\{0,0\})$ & $J_{1,\mu\nu}^{A(S)}$& $1^{++}$ & $4.70^{+0.12}_{-0.11}$ & $27(\pm5\%)$ & $3.27\sim3.92$ &27.3& $4.91^{+0.11}_{-0.12}$ & $28(\pm5\%)$ & $3.56\sim4.20$ &26.5\\ 
  $(2,0\{0,0\})$ & $J_{7,\mu\nu}^{A(S)}$& $1^{+-}$ & $4.78^{+0.12}_{-0.11}$ & $27(\pm5\%)$ & $3.58\sim4.16$ &25.4& $4.89^{+0.10}_{-0.11}$ & $28(\pm5\%)$ & $3.60\sim4.50$ &28.5\\ 
  $(1,1\{1,0\})$ & $J_{2,\mu\nu}^{A(S)}$& $1^{++}$ & $4.80^{+0.12}_{-0.16}$ & $28(\pm5\%)$ & $3.15\sim3.94$ &39.6& $4.84^{+0.12}_{-0.16}$ & $29(\pm5\%)$ & $2.63\sim4.13$ &37.9\\ 
  $(1,1\{1,0\})$ & $J_{8,\mu\nu}^{A(S)}$& $1^{+-}$ & $4.81\pm0.10$ & $27(\pm5\%)$ & $3.71\sim4.51$ &26.3& $4.85^{+0.11}_{-0.10}$ & $28(\pm5\%)$ & $4.69\sim5.16$ &28.2\\ 
  $(1,1\{0,1\})$ & $J_{3,\mu\nu}^{A(S)}$& $1^{++}$ & $4.80^{+0.11}_{-0.10}$ & $26(\pm5\%)$ & $2.75\sim3.31$ &26.1& $4.82^{+0.12}_{-0.11}$ & $27(\pm5\%)$ & $3.37\sim 4.11$ &47.0\\ 
  $(1,1\{0,1\})$ & $J_{9,\mu\nu}^{A(S)}$& $1^{+-}$ & $4.98^{+0.13}_{-0.23}$ & $26(\pm5\%)$ & $2.73\sim3.14$ &24.0& $4.92^{+0.11}_{-0.10}$ & $28(\pm5\%)$ & $3.55\sim3.91$ &23.4\\ 
  $(0,2\{1,1\})$ & $J_{4,\mu\nu}^{A(S)}$& $1^{++}$ & $4.80^{+0.10}_{-0.11}$ & $26(\pm5\%)$ & $2.51\sim3.14$ &27.5& $4.80^{+0.10}_{-0.11}$ & $26(\pm5\%)$ & $2.52\sim3.15$ &27.4\\ 
  $(0,2\{1,1\})$ & $J_{10,\mu\nu}^{A(S)}$& $1^{+-}$ & $4.83^{+0.10}_{-0.11}$ & $28(\pm5\%)$ & $3.06\sim3.82$ &28.6& $4.83^{+0.10}_{-0.12}$ & $28(\pm5\%)$ & $3.08\sim3.82$ &28.3\\ 
  $(0,2\{2,0\})$ & $J_{5,\mu\nu}^{A(S)}$& $1^{++}$ & $5.16^{+0.12}_{-0.13}$ & $30(\pm5\%)$ & $2.94\sim3.90$ &41.4& $4.69\pm0.09$ & $24(\pm5\%)$ & $2.22\sim2.82$ &27.5\\
  $(0,2\{2,0\})$ & $J_{11,\mu\nu}^{A(S)}$& $1^{+-}$ & $5.19^{+0.12}_{-0.13}$ & $30(\pm5\%)$ & $3.55\sim3.92$ &43.4& $4.67\pm0.09$ & $23(\pm5\%)$ & $2.69\sim2.87$ &21.6\\ 
  $(1,1)_\mathrm{mix}$ & $J_{2,\mu\nu}^{A(S)}+J_{3,\mu\nu}^{A(S)}$& $1^{++}$ & $4.80\pm0.10$ & $27(\pm5\%)$ & $3.01\sim3.76$ &24.1& $4.93^{+0.09}_{-0.10}$ & $29(\pm5\%)$ & $3.22\sim4.02$ &38.4\\ 
  $(1,1)_\mathrm{mix}$ & $J_{8,\mu\nu}^{A(S)}+J_{9,\mu\nu}^{A(S)}$& $1^{+-}$ & $4.80^{+0.11}_{-0.13}$ & $26(\pm5\%)$ & $2.71\sim3.13$ &30.2& $4.94\pm0.10$ & $29(\pm5\%)$ & $3.37\sim4.21$ &38.2\\ 
  \hline
  \hline
\end{tabular}
}
\end{table}

\section{Conclusion and Discussion}\label{Resultanddis}
We have investigated the mass spectra for the D-wave $cs\bar{c}\bar{s}$ tetraquark states with $J^{PC}=1^{++}$ and  $1^{+-}$ in the framework of QCD sum rules. We construct the D-wave non-local interpolating tetraquark currents with  covariant derivative operators in the $(L_\lambda,L_\rho\{l_{\rho_1},l_{\rho_2}\})=(2,0\{0,0\}),(1,1\{1,0\}),(1,1\{0,1\}),(0,2\{1,1\}),(0,2\{2,0\}),(0,2\{0,2\})$ excitation structures. The two-point correlation functions are calculated up to dimension ten condensates in the leading order of $\alpha_s$. We establish reliable mass sum rules for all these currents and obtain  the mass spectra of D-wave $cs\bar{c}\bar{s}$ tetraquarks, as shown in Table~\ref{cscsResultTab}. Our results support to interpret the recently observed $X(4685)$ structure as a D-wave $cs\bar{c}\bar{s}$ tetraquark state with $J^{PC}=1^{++}$ in the $(2,0\{0,0\})$ or $(0,2\{2,0\})$ excitation mode. However, some other possibilities of the excitation modes cannot be excluded by our results within errors.


The mass spectra of $cs\bar c\bar s$ tetraquark states in different color configurations have been studied in Ref.~\cite{Deng:2019dbg}, in which the masses of color symmetric tetraquarks are lower than the color antisymmetric tetraquarks in ground state ($L=0$). Similar results were also obtained for the fully heavy tetraquark states~\cite{Deng:2020iqw,Wang:2021kfv,Yang:2021zrc}. However,  the situation is different for the excited $cs\bar c\bar s$ tetraquarks that the masses of color antisymmetric tetraquarks are lower than the color symmetric tetraquarks. Such behavior is also consistent with our results in Table~\ref{cscsResultTab} for the D-wave $cs\bar c\bar s$ tetraquarks, except for those in the $(0,2\{2,0\})$ structures with two $\rho$-mode excitations. In Table~\ref{cscsResultTab}, the masses for the positive $\mathbb{C}$-parity tetraquarks provide relation $6_{\rho\rho}<3_{\lambda\lambda}<3_{\lambda\rho}<3_{\rho\rho}$ and negative $\mathbb{C}$-parity tetraquarks with $6_{\rho\rho}<3_{\lambda\lambda}<6_{\lambda\lambda}<3_{\rho\rho}$, which is consistent with the conclusion for the P-wave $cc\bar c\bar c$ systems~\cite{Wang:2021kfv}. 

\begin{figure}[t!]
\centering
\includegraphics[width=10cm]{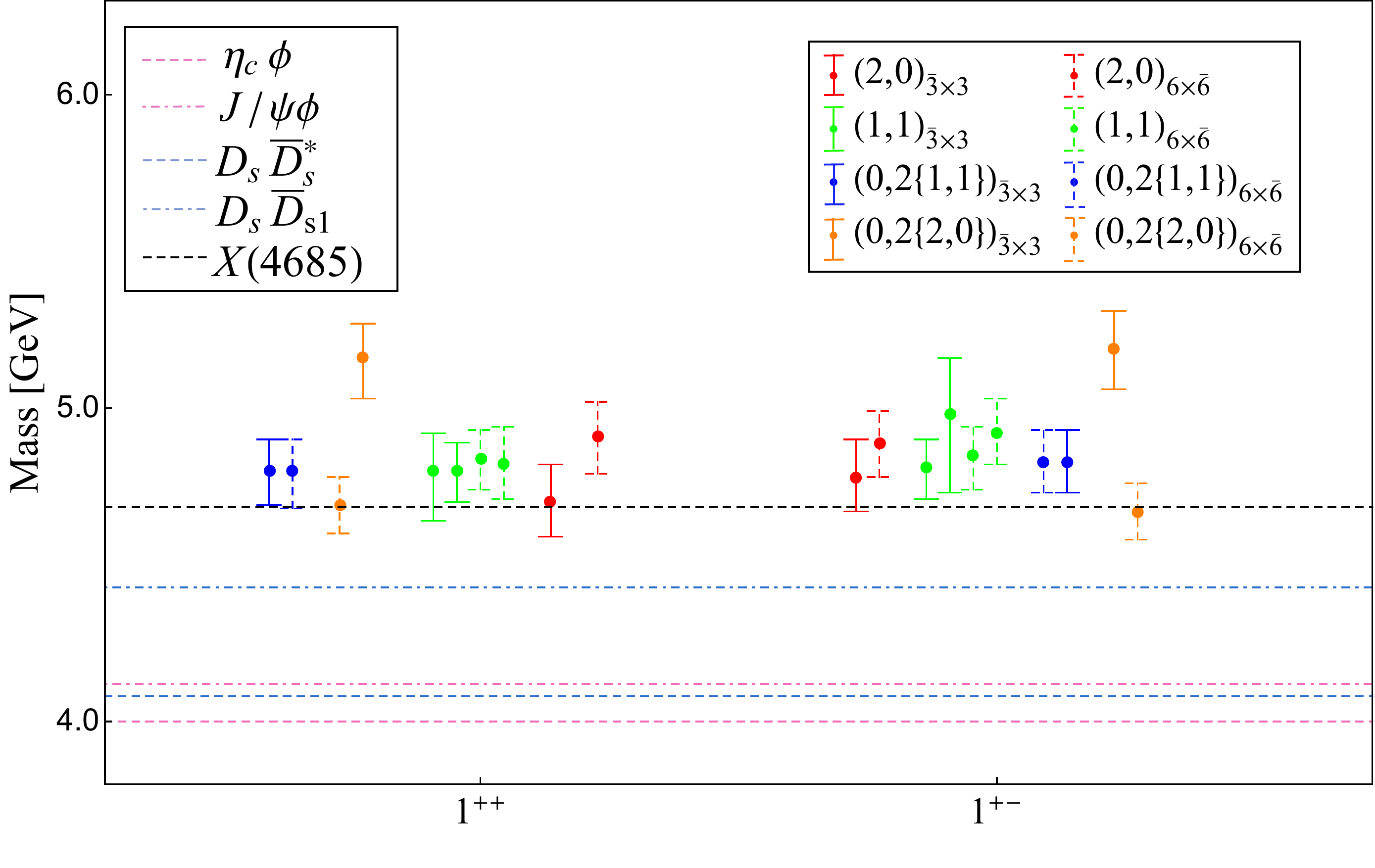}\\
\caption{Mass spectra for D-wave $cs\bar{c}\bar{s}$ tetraquark with $J^{PC}=1^{++}$ and $1^{+-}$, comparing with the corresponding two-meson mass thresholds.}
\label{FigcscsSpectrum}
\end{figure}
We collect the mass spectra of these $cs\bar{c}\bar{s}$ tetraquarks comparing with the corresponding two-meson open-charm mass thresholds in Fig.~\ref{FigcscsSpectrum}. It is clearly that these D-wave $cs\bar{c}\bar{s}$ tetraquarks with $J^{PC}=1^{++}$ and $1^{+-}$ lie above the mass thresholds of $D_s\bar{D}_s^*$, $J/\psi\phi$ and $\eta_c\phi$. Accordingly, we list their possible decay channels in both S-wave and P-wave in Table~\ref{QsQsDecayTab}. We suggest searching for these D-wave $cs\bar{c}\bar{s}$ tetraquarks in both the hidden-charm channels $J/\psi\phi$, $\eta_c\phi$ and open-charm channels such as $D_s\bar{D}_s^*$, $D_{s}\bar{D}_{s1}^*$ and so on. 

\begin{table}[h!]
\caption{Possible decay channels of the D-wave $cs\bar{c}\bar{s}$ tetraquark states with $J^{PC}=1^{++}$ and $1^{+-}$.}\label{QsQsDecayTab}
\renewcommand\arraystretch{1.3} 
\setlength{\tabcolsep}{2.em}{ 
\begin{tabular}{c c c }
  \hline
   \hline
$J^{PC}$ & $S$-wave & $P$-wave \\
  \hline
 $1^{++}$ & $D_{s0}^*\bar{D}_{s1},D_{s}\bar{D}_s^*,D_{s}\bar{D}_{s1}^*,$ & $D_s\bar{D}_{s1},D_{s0}^*\bar{D}_{s}^*,D_{s0}^*\bar{D}_{s1}^*,$ \vspace{1ex}\\
                & $D_{s1}\bar{D}_{s1},D_{s1}\bar{D}_{s2}^*,$ & $D_{s}^*\bar{D}_{s1},D_{s1}^*\bar{D}_{s1},D_s^*\bar{D}_{s2}^*,$ \vspace{1ex}\\
                & $J/\psi\phi$ & $D_{s1}^*\bar{D}_{s2}^*,h_c(1P)\phi$ \vspace{1ex}\\  
 $1^{+-}$ & $D_{s0}^*\bar{D}_{s1},D_{s}\bar{D}_s^*,D_{s}\bar{D}_{s1}^*,$ & $D_s\bar{D}_{s1},D_{s0}^*\bar{D}_{s}^*,D_{s0}^*\bar{D}_{s1}^*,$\vspace{1ex}\\    
               & $D_{s1}\bar{D}_{s1},D_{s1}\bar{D}_{s2}^*,$ & $D_{s}^*\bar{D}_{s1},D_{s1}^*\bar{D}_{s1},D_s^*\bar{D}_{s2}^*,$\vspace{1ex}\\    
               & $\eta_c\phi$ & $\chi_{c0}(1P)\phi,\chi_{c1}(1P)\phi$ \vspace{1ex}\\   
  \hline
   \hline
\end{tabular}
}
\end{table}

\section*{ACKNOWLEDGMENTS}
This work is supported by the National Natural Science Foundation of China under Grant No. 12175318, the National Key R$\&$D Program of China under Contracts No. 2020YFA0406400, the Natural Science Foundation of Guangdong Province of China under Grant No. 2022A1515011922, and the Fundamental Research Funds for the Central Universities.

\section*{Appendix: Spectrum function for D-wave interpolating current \label{appendix}} 
\par The spectral functions for D-wave interpolating current $J_{i}^{A(S)}$ can be written as
\begin{eqnarray}
\nonumber\rho_{i;A(S)}(s)&=&\rho^{pert}_{i;A(S)}(s)+\langle \bar{s}s\rangle\rho^{\langle \bar{s}s\rangle}_{i;A(S)}(s)+m_s\langle\bar{s}s\rangle\rho^{m_s\langle \bar{s}s\rangle}_{i;A(S)}(s)+\langle g_s^2G^2\rangle\rho^{\langle g_s^2G^2\rangle}_{i;A(S)}(s)+\langle \bar{s}\sigma\cdot G s\rangle\rho^{\langle \bar{s}\sigma\cdot G s\rangle}_{i;A(S)}(s)+m_s\langle \bar{s}\sigma\cdot G s\rangle\rho^{m_s\langle \bar{s}\sigma\cdot G s\rangle}_{i;A(S)}(s)\\
\nonumber& &+\langle \bar{s}s \bar{s}s\rangle\rho^{\langle \bar{s}s \bar{s}s\rangle}_{i;A(S)}(s)+\langle \bar{s}s\rangle\langle \bar{s}\sigma\cdot G s\rangle\rho^{\langle \bar{s}s\rangle\langle \bar{s}\sigma\cdot G s\rangle}_{i;A(S)}(s)+\langle g_s^2G^2\rangle\langle \bar{s}s\rangle\rho^{\langle g_s^2G^2\rangle\langle \bar{s}s\rangle}_{i;A(S)}(s)+m_s\langle g_s^2G^2\rangle\langle \bar{s}s\rangle\rho^{m_s\langle g_s^2G^2\rangle\langle \bar{s}s\rangle}_{i;A(S)}(s)\\
& &+\langle g_s^2G^2\rangle^2\rho^{\langle g_s^2G^2\rangle^2}_{i;A(S)}(s)+\langle g_s^2G^2\rangle\langle \bar{s}\sigma\cdot G s\rangle\rho^{\langle g_s^2G^2\rangle\langle \bar{s}\sigma\cdot G s\rangle}_{i;A(S)}(s)+m_s\langle g_s^2G^2\rangle\langle \bar{s}\sigma\cdot G s\rangle\rho^{m_s\langle g_s^2G^2\rangle\langle \bar{s}\sigma\cdot G s\rangle}_{i;A(S)}(s).
\end{eqnarray}
The spectral functions for $(2,0\{0,0\})$ structure are shown as following:
{\allowdisplaybreaks

}
where $F(s,x,y)=\frac{m_c^2 (1-x y)}{1-x}-s (1-y) y$, $G(s,z)=m_c^2-s(1-z)z$, $y_{max}=\frac{1}{2}+\frac{\sqrt{4 m_c^2 s (x-1)+\left(s (x-1)-m_c^2 x\right)^2}+m_c^2 x}{2 s (1-x)}$, $y_{min}=\frac{1}{2}-\frac{\sqrt{4 m_c^2 s (x-1)+\left(s (x-1)-m_c^2 x\right)^2}-m_c^2 x}{2 s (1-x)}$, $x_{max}=\left(1-2 \sqrt{m_c^2/s}\right)/\left(\sqrt{m_c^2/s}-1\right)^2$, $z_{max}=\frac{1}{2}\left(1+\sqrt{1-4 m_c^2/s}\right)$, $z_{min}=\frac{1}{2}\left(1-\sqrt{1-4 m_c^2/s}\right)$, coefficient $c_p=1$ for current $J_{1,\mu\nu}^{A(S)}$ while $c_p=-1$ for current $J_{7,\mu\nu}^{A(S)}$, and $c_1=12,c_2=-8,c_3=4$ for color antisymmetric current $J_{i,\mu\nu}^{A}$ while $c_1=24,c_2=8,c_3=20$ for color symmetric current $J_{i,\mu\nu}^{S}$. 
The spectral functions for $(1,1\{1,0\})$ structure are shown as:
{\allowdisplaybreaks

}
where the coefficient $c_p=1$ for current $J_{2,\mu\nu}^{A(S)}$  while $c_p=-1$ for current $J_{8,\mu\nu}^{A(S)}$.
The spectral functions for $(1,1\{0,1\})$ structure are shown as:
{\allowdisplaybreaks
%
}
where the coefficient $c_p=1$ for current $J_{3,\mu\nu}^{A(S)}$  while $c_p=-1$ for current $J_{9,\mu\nu}^{A(S)}$.
The spectral functions for $(0,2\{1,1\})$ structure with $\mathbb{C}=+1$ are shown as 
{\allowdisplaybreaks
\begin{eqnarray}
\nonumber\rho^{pert}_{4;A(S)}(s)&=&\int^{x_{max}}_{0}dx\int^{y_{max}}_{y_{min}}dy\frac{c_1 x^3}{12902400 \pi ^5 (y-1)^5} \left(c_p+1\right) F(s,x,y)^3 \left(-15 (y-1) F(s,x,y) \left(7 s x y m_c m_s (11 x y (2 x y+3)\right.\right.\\
\nonumber& &\left.-10) (x y-1)+14 m_c^2 m_s^2 (x y (2 x y-5)+5)-6 s^2 (x-1) x^2 (y-1) y^3 (5 x y (5 x y+7)-7)\right)+21 x F(s,x,y)^2 \Big(20 s\\
\nonumber& &(x-1) x (y-1) y^2 (x y+2) (3 x y-1)-m_c m_s (x y-1) (x y+3) (8 x y-5)\Big)+2 (x-1) x^2 y (10 x y (3 x y+7)-49)\\
\nonumber& &F(s,x,y)^3-60 s x (y-1)^2 y^2 \Big(7 s x y m_c m_s (x y-1) (4 x y+5)+14 m_c^2 m_s^2 (4 x y-5)-2 s^2 (x-1) x^2 (y-1) y^3 (6 x y\\
\nonumber& &+7)\Big)\Big),\\
\nonumber\rho^{\langle\bar{s}s\rangle}_{4;A(S)}(s)&=&0,\\
\nonumber\rho^{\langle m_s\bar{s}s\rangle}_{4;A(S)}(s)&=&0,\\
\nonumber\rho^{\langle g_sG^2\rangle}_{4;A(S)}(s)&=&\int^{x_{max}}_{0}dx\int^{y_{max}}_{y_{min}}dy\Bigg\{\frac{c_1 x^3}{15482880 \pi ^5 (x-1)^3 (y-1)^5} m_c \left(c_p+1\right) \left(6 (y-1) F(s,x,y) \left(6 s (x-1) x (y-1) y^2 m_c \Big(s x y\right.\right.\\
\nonumber& &(x y-1) (5 x y (5 x y+7)-7) (y (((x-3) x+3) y+x-3)+1)-7 m_s^2 (4 x y-5) (y (((x-2) x+2) y-2)+1)\Big)\\
\nonumber& &-7 s x y m_c^2 m_s (x y-1)^2 (11 x y (2 x y+3)-10) (y (((x-3) x+3) y+x-3)+1)-14 m_c^3 m_s^2 (x y-1) (x y (2 x y-5)\\
\nonumber& &+5) (y (((x-3) x+3) y+x-3)+1)-21 s^2 (x-1) x^2 (y-1) y^3 m_s (x y-1) (4 x y+5) (y (((x-3) x+3) y+x-3)\\
\nonumber& &+1)\Big)-21 F(s,x,y)^2 \left(2 (x-1) (y-1) m_c \left(3 m_s^2 (x y (2 x y-5)+5) (y (((x-2) x+2) y-2)+1)-10 s x^2 y^2 (x y-1)\right.\right.\\
\nonumber& &\left.(x y+2) (3 x y-1) (y (((x-3) x+3) y+x-3)+1)\right)+x m_c^2 m_s (8 x y-5) (x y-1)^2 (y (((x-3) x+3) y+x-3)\\
\nonumber& &\left.+1) (x y+3)+3 s (x-1) x (y-1) y m_s (11 x y (2 x y+3)-10) (x y-1) (y (((x-3) x+3) y+x-3)+1)\right)+(x-1)\\
\nonumber& &x (y (y (x ((x-3) x+3) y-3)+3)-1) F(s,x,y)^3 \left(4 x y m_c (10 x y (3 x y+7)-49)-21 m_s (x y+3) (8 x y-5)\right)\\
\nonumber& &-6 s x (y-1)^2 y^2 m_c (y (y (x ((x-3) x+3) y-3)+3)-1) \left(7 s x y m_c m_s (x y-1) (4 x y+5)+14 m_c^2 m_s^2 (4 x y-5)\right.\\
\nonumber& &\left.\left.-2 s^2 (x-1) x^2 (y-1) y^3 (6 x y+7)\right)\right)+\frac{c_3}{11796480 \pi ^5 (y-1)^3} (x-1) x^2 \left(c_p+1\right) F(s,x,y)^2 \left(4 s (y-1) y (293 x y\right.\\
\nonumber& &\left.-27) F(s,x,y)+(181 x y-36) F(s,x,y)^2+588 s^2 x (y-1)^2 y^3\right)\Bigg\},\\
\nonumber\rho^{\langle\bar{s}\sigma\cdot Gs\rangle}_{4;A(S)}(s)&=&\int^{x_{max}}_{0}dx\int^{y_{max}}_{y_{min}}dy\Bigg\{-\frac{c_1 x^2}{512 \pi ^3 (y-1)^3} m_c \left(c_p+1\right) (x y-1) F(s,x,y) \left(s (y-1) y (5 x y-2) F(s,x,y)+(x y-1)\right.\\
\nonumber& &\left. F(s,x,y)^2+2 s^2 x (y-1)^2 y^3\right)\Bigg\},\\
\nonumber\rho^{m_s\langle\bar{s}\sigma\cdot Gs\rangle}_{4;A(S)}(s)&=&-\int^{x_{max}}_{0}dx\int^{y_{max}}_{y_{min}}dy\Bigg\{-\frac{c_1 x}{2304 \pi ^3 (y-1)^2} \left(c_p+1\right) \left(18 s x (y-1) y^2 F(s,x,y) \left(s (x-1) (y-1) y (5 x y-1)-m_c^2\right)\right.\\
\nonumber& &\left.+3 F(s,x,y)^2 \left(m_c^2 (3-3 x y)+10 s (x-1) x (y-1) y^2 (5 x y-2)\right)+2 (x-1) x y (10 x y-7) F(s,x,y)^3+6 s^3 (x-1)\right.\\
\nonumber& &\left. x^2 (y-1)^3 y^5\right)\Bigg\},\\
\nonumber\rho^{\langle\bar{s}s\bar{s}s\rangle}_{4;A(S)}(s)&=&0,\\
\nonumber\rho^{\langle\bar{s}s\rangle\langle\bar{s}\sigma\cdot Gs\rangle}_{4;A(S)}(s)&=&0,\\
\nonumber\rho^{\langle g_sG^2\rangle\langle \bar{s}s\rangle}_{4;A(S)}(s)&=&0,\\
\nonumber\rho^{m_s\langle g_sG^2\rangle\langle \bar{s}s\rangle}_{4;A(S)}(s)&=&0,\\
\nonumber\rho^{\langle g_sG^2\rangle^2\;A(S)}_4(s)&=&\int^{x_{max}}_{0}dx\int^{y_{max}}_{y_{min}}dy\frac{c_1 x^5 y^4 m_c^4 \left(c_p+1\right) (10 x y (3 x y+7)-49)}{46448640 \pi ^5 (x-1)^2 (y-1)^2}\\
\nonumber& &+\int^{1}_{0}dx\int^{1}_{0}dy\frac{c_1 s x^5 y^5 m_c^4 \left(c_p+1\right) (x y+2) (3 x y-1)}{1327104 \pi ^5 (x-1)^2 (y-1)},\\
\nonumber\rho^{\langle g_sG^2\rangle\langle \bar{s}\sigma\cdot Gs\rangle}_{4;A(S)}(s)&=&\int^{x_{max}}_{0}dx\int^{y_{max}}_{y_{min}}dy\Bigg\{-\frac{c_1 x^2}{6144 \pi ^3 (x-1)^3 (y-1)^3} m_c \left(c_p+1\right) (y (y (x ((x-3) x+3) y-3)+3)-1) \Big(3 (x-1)\\
\nonumber& & (x y-1) F(s,x,y)+m_c^2 (x y-1)^2+s (x-1) (y-1) y (5 x y-2)\Big)\Bigg\}+\int^{1}_{0}dx\int^{1}_{0}dy\Bigg\{-\frac{c_1 s x^2 y}{18432 \pi ^3 (x-1)^3 (y-1)^2}\\
\nonumber& & m_c \left(c_p+1\right) (y (y (x ((x-3) x+3) y-3)+3)-1) \left(m_c^2 (5 x y-2) (x y-1)+3 s (x-1) x (y-1) y^2\right)\Bigg\},\\
\nonumber\rho^{m_s\langle g_sG^2\rangle\langle \bar{s}\sigma\cdot Gs\rangle}_{10;A(S)}(s)&=&\int^{x_{max}}_{0}dx\int^{y_{max}}_{y_{min}}dy\Bigg\{-\frac{c_1 x }{27648 \pi ^3 (x-1)^2 (y-1)^2}m_c^2 \left(c_p+1\right) (x y-1)^2 \left(x y \left(20 ((x-3) x+3) y^2+26 (x-3) y\right.\right.\\
\nonumber& &\left.+23\Big)+18 (y-1) y+9\right)\Bigg\}+\int^{1}_{0}dx\int^{1}_{0}dy\Bigg\{-\frac{c_1 x}{27648 \pi ^3 (x-1)^3 (y-1)^2} m_c^2 \left(c_p+1\right) \left(s (x-1) x (y-1) y^2 (x y (x\right.\\
\nonumber& & y(10 y (5 ((x-3) x+3) y-2 x+6)-9)+42 (4-5 y) y-50)+42 (y-1) y+11)-3 m_c^2 (x y-1)^2 (y (((x-3) x\\
\nonumber& &\left.+3) y+x-3)+1)\right)\Bigg\},\\
\end{eqnarray}
}
where the coefficient $c_p=1$. The spectral functions for $(0,2\{1,1\})$ structure with $\mathbb{C}=-1$ are shown as 
{\allowdisplaybreaks
\begin{eqnarray}
\nonumber\rho^{pert}_{10;A(S)}(s)&=&-\int^{x_{max}}_{0}dx\int^{y_{max}}_{y_{min}}dy\frac{c_1 x^3}{614400 \pi ^5 (y-1)^5} \left(c_p-1\right) F(s,x,y)^3 \left(5 (y-1) F(s,x,y) \left(2 s^2 (x-1) x^2 (y-1) y^3 (11 x y\right.\right.\\
\nonumber& &\left.-3)-5 m_c m_s (x y-1) \left(s x y (5 x y-2)-2 m_c m_s\right)\right)+x F(s,x,y)^2 \left(8 s (x-1) x (y-1) y^2 (7 x y-4)-15 m_c m_s\right.\\
\nonumber& &\left.(x y-1)^2\right)+4 (x-1) x^2 y (x y-1) F(s,x,y)^3+20 s x (y-1)^2 y^2 \left(5 m_c m_s \left(2 m_c m_s+s x y (1-x y)\right)+2 s^2 (x-1)\right.\\
\nonumber& &\left.\left.x^2 (y-1) y^3\right)\right),\\
\nonumber\rho^{\langle\bar{s}s\rangle}_{10;A(S)}(s)&=&0,\\
\nonumber\rho^{\langle m_s\bar{s}s\rangle}_{10;A(S)}(s)&=&0,\\
\nonumber\rho^{\langle g_sG^2\rangle}_{10;A(S)}(s)&=&\int^{x_{max}}_{0}dx\int^{y_{max}}_{y_{min}}dy\Bigg\{-\frac{c_1 x^3}{737280 \pi ^5 (x-1)^3 (y-1)^5} m_c \left(c_p-1\right) \left(2 (y-1) F(s,x,y) \left(2 s (x-1) x (y-1) y^2 m_c\right.\right. \\
\nonumber& &\left(15 m_s^2 (y (((x-2) x+2) y-2)+1)+s x y (11 x y-3) (x y-1) (y (((x-3) x+3) y+x-3)+1)\right)-5 s x y m_c^2 m_s\\
\nonumber& &(x y-1)^2 (5 x y-2) (y (((x-3) x+3) y+x-3)+1)+10 m_c^3 m_s^2 (x y-1)^2 (y (((x-3) x+3) y+x-3)+1)\\
\nonumber& &\left.-15 s^2 (x-1) x^2 (y-1) y^3 m_s (y (y (x ((x-3) x+3) y-3)+3)-1)\right)+(x y-1) F(s,x,y)^2 \left(2 (x-1) (y-1) m_c\right.\\
\nonumber& &\left(15 m_s^2 (y (((x-2) x+2) y-2)+1)+4 s x^2 y^2 (7 x y-4) (y (((x-3) x+3) y+x-3)+1)\right)-15 x m_c^2 m_s (x y-1)^2\\
\nonumber& &\left.(y (((x-3) x+3) y+x-3)+1)-15 s (x-1) x (y-1) y m_s (5 x y-2) (y (((x-3) x+3) y+x-3)+1)\right)\\
\nonumber& &+(x-1) x (x y-1)^2 (y (((x-3) x+3) y+x-3)+1) F(s,x,y)^3 \left(8 x y m_c-15 m_s\right)+2 s x (y-1)^2 y^2 m_c (y (y (x ((x\\
\nonumber& &\left.-3) x+3) y-3)+3)-1) \left(-5 s x y m_c m_s (x y-1)+10 m_c^2 m_s^2+2 s^2 (x-1) x^2 (y-1) y^3\right)\right)+\frac{c_3}{2359296 \pi ^5 (y-1)^3}\\
\nonumber& & (x-1) x^2 \left(c_p-1\right) F(s,x,y)^2 \left(4 s (y-1) y (15 x y-2) F(s,x,y)+(9 x y-3) F(s,x,y)^2+36 s^2 x (y-1)^2 y^3\right)\Bigg\},\\
\nonumber\rho^{\langle\bar{s}\sigma\cdot Gs\rangle}_{10;A(S)}(s)&=&\int^{x_{max}}_{0}dx\int^{y_{max}}_{y_{min}}dy\Bigg\{\frac{c_1 x^2}{512 \pi ^3 (y-1)^3} m_c \left(c_p-1\right) (x y-1) F(s,x,y) \left(s (y-1) y (5 x y-2) F(s,x,y)+(x y-1)\right.\\
\nonumber& &\left. F(s,x,y)^2+2 s^2 x (y-1)^2 y^3\right)\Bigg\},\\
\nonumber\rho^{m_s\langle\bar{s}\sigma\cdot Gs\rangle}_{10;A(S)}(s)&=&\int^{x_{max}}_{0}dx\int^{y_{max}}_{y_{min}}dy\Bigg\{\frac{c_1 x}{768 \pi ^3 (y-1)^2} \left(c_p-1\right) \left(2 s x (y-1) y^2 F(s,x,y) \left(s (x-1) (y-1) y (11 x y-3)-3 m_c^2\right)\right.\\
\nonumber& &+F(s,x,y)^2 \left(m_c^2 (3-3 x y)+4 s (x-1) x (y-1) y^2 (7 x y-4)\right)+4 (x-1) x y (x y-1) F(s,x,y)^3+2 s^3 (x-1) x^2\\
\nonumber& &\left. (y-1)^3 y^5\right)\Bigg\},\\
\nonumber\rho^{\langle\bar{s}s\bar{s}s\rangle}_{10;A(S)}(s)&=&0,\\
\nonumber\rho^{\langle\bar{s}s\rangle\langle\bar{s}\sigma\cdot Gs\rangle}_{10;A(S)}(s)&=&0,\\
\nonumber\rho^{\langle g_sG^2\rangle\langle \bar{s}s\rangle}_{10;A(S)}(s)&=&0,\\
\nonumber\rho^{m_s\langle g_sG^2\rangle\langle \bar{s}s\rangle}_{10;A(S)}(s)&=&0,\\
\nonumber\rho^{\langle g_sG^2\rangle^2}_{10;A(S)}(s)&=&-\int^{x_{max}}_{0}dx\int^{y_{max}}_{y_{min}}dy\frac{c_1 x^5 y^4 m_c^4 \left(c_p-1\right) (x y-1)}{1105920 \pi ^5 (x-1)^2 (y-1)^2}-\int^{1}_{0}dx\int^{1}_{0}dy\frac{i c_1 s x^5 y^5 m_c^4 \left(c_p-1\right) (7 x y-4)}{3317760 \pi ^5 (x-1)^2 (y-1)},\\
\nonumber\rho^{\langle g_sG^2\rangle\langle \bar{s}\sigma\cdot Gs\rangle}_{10;A(S)}(s)&=&\int^{x_{max}}_{0}dx\int^{y_{max}}_{y_{min}}dy\Bigg\{\frac{c_1 x^2}{6144 \pi ^3 (x-1)^3 (y-1)^3} m_c \left(c_p-1\right) (y (y (x ((x-3) x+3) y-3)+3)-1) \left(3 (x-1) (x y\right.\\
\nonumber& &\left.-1) F(s,x,y)+m_c^2 (x y-1)^2+s (x-1) (y-1) y (5 x y-2)\right)\Bigg\}+\int^{1}_{0}dx\int^{1}_{0}dy\Bigg\{\frac{c_1 s x^2 y}{18432 \pi ^3 (x-1)^3 (y-1)^2} m_c\\
\nonumber& & \left(c_p-1\right) (y (y (x ((x-3) x+3) y-3)+3)-1) \left(m_c^2 (5 x y-2) (x y-1)+3 s (x-1) x (y-1) y^2\right)\Bigg\},\\
\nonumber\rho^{m_s\langle g_sG^2\rangle\langle \bar{s}\sigma\cdot Gs\rangle}_{5,11;A(S)}(s)&=&\int^{x_{max}}_{0}dx\int^{y_{max}}_{y_{min}}dy\Bigg\{\frac{c_1 x}{9216 \pi ^3 (x-1)^2 (y-1)^2} m_c^2 \left(c_p-1\right) (x y-1) \left(y \left(x \left(y \Big(x \left(4 ((x-3) x+3) y^2-3\right)-12 y\right.\right.\right.\\
\nonumber& &\left.\left.\left.+18\Big)-4\right)-6 y+6\right)-3\right)\Bigg\}+\int^{1}_{0}dx\int^{1}_{0}dy\Bigg\{\frac{c_1 x}{27648 \pi ^3 (x-1)^3 (y-1)^2} m_c^2 \left(c_p-1\right) \left(s (x-1) x (y-1) y^2\right.\\
\nonumber& &\left(x y \left(y \left(28 ((x-3) x+3) x y^2-4 (4 (x-3) x+33) y-9 x+102\right)-28\right)+30 (y-1) y+7\right)-3 m_c^2 (x y-1)^2\\
\nonumber& &\left.(y (((x-3) x+3) y+x-3)+1)\right)\Bigg\},\\
\end{eqnarray}
}
where the coefficient $c_p=-1$. 
The spectral functions for $(0,2\{2,0\})$ structure are shown as 
{\allowdisplaybreaks
\begin{eqnarray}
\nonumber\rho^{pert}_{5,11;A(S)}(s)&=&-\int^{x_{max}}_{0}dx\int^{y_{max}}_{y_{min}}dy\frac{c_1 x^3}{6451200 \pi ^5 (y-1)^5} F(s,x,y)^3 \left(15 (y-1) F(s,x,y) \left(7 s x y m_c m_s (x y-1) \left(x y \left(11 x y \left(c_p+6\right)\right.\right.\right.\right.\\
\nonumber& &\left.\left.+4 c_p+99\right)-25\right)+14 m_c^2 m_s^2 \left(x y \left(x y \left(c_p+6\right)-15\right)+10\right)-s^2 (x-1) x^2 (y-1) y^3 \left(x y \left(25 x y \left(3 c_p+10\right)+28\right.\right.\\
\nonumber& &\left.\left.\left.\left(c_p+13\right)\right)-63\right)\right)+21 x F(s,x,y)^2 \left(m_c m_s (x y-1) \left(x y \left(4 x y \left(c_p+6\right)+2 c_p+57\right)-35\right)-2 s (x-1) x (y-1) y^2\right.\\
\nonumber& &\left.\left(5 x^2 y^2 \left(3 c_p+10\right)+x y \left(11 c_p+78\right)-2 \left(c_p+12\right)\right)\right)-(x-1) x^2 y \left(10 x^2 y^2 \left(3 c_p+10\right)+28 x y \left(c_p+8\right)\right.\\
\nonumber& &\left.-7 \left(c_p+18\right)\right) F(s,x,y)^3+60 s x (y-1)^2 y^2 \left(7 s x y m_c m_s (x y-1) \left(2 x y \left(c_p+6\right)+15\right)+14 m_c^2 m_s^2 \left(2 x y \left(c_p+6\right)\right.\right.\\
\nonumber& &\left.\left.\left.-15\right)-2 s^2 (x-1) x^2 (y-1) y^3 \left(x y \left(3 c_p+10\right)+14\right)\right)\right),\\
\nonumber\rho^{\langle\bar{s}s\rangle}_{5,11;A(S)}(s)&=&\int^{x_{max}}_{0}dx\int^{y_{max}}_{y_{min}}dy\frac{c_1 x^3}{384 \pi ^3 (y-1)^4} m_c (x y-1) F(s,x,y)^2 \Big(s (y-1) y (x y (11 x y+14)-3) F(s,x,y)+(x y (x y+2)\\
\nonumber& &-1) F(s,x,y)^2+6 s^2 x (y-1)^2 y^3 (x y+1)\Big),\\
\nonumber\rho^{\langle m_s\bar{s}s\rangle}_{5,11;A(S)}(s)&=&-\int^{x_{max}}_{0}dx\int^{y_{max}}_{y_{min}}dy\frac{c_1 x^2}{384 \pi ^3 (y-1)^3} F(s,x,y) \left(-3 s x (y-1) y^2 F(s,x,y) \left(4 m_c^2 (x y-1)+s (x-1) (y-1) y (x y (25\right.\right.\\
\nonumber& &\left.x y+26)-3)\right)-2 F(s,x,y)^2 \left(m_c^2 (x y-1)^2+s (x-1) x (y-1) y^2 (x y (35 x y+39)-8)\right)-(x-1) x y (x y (5 x y+8)\\
\nonumber& &\left.-3) F(s,x,y)^3-12 s^3 (x-1) x^2 (y-1)^3 y^5 (x y+1)\right),\\
\nonumber\rho^{\langle g_sG^2\rangle}_{5,11;A(S)}(s)&=&\int^{x_{max}}_{0}dx\int^{y_{max}}_{y_{min}}dy\Bigg\{-\frac{c_1 x^3}{7741440 \pi ^5 (x-1)^3 (y-1)^5} m_c \left(-6 (y-1) F(s,x,y) \left(-21 s^2 (x-1) x^2 (y-1) y^3 m_s (y (y (x\right.\right.\\
\nonumber& &((x-3) x+3) y-3)+3)-1) \left(2 x y \left(c_p+6\right)+15\right)+s (x-1) x (y-1) y^2 m_c \left(s x y (y (y (x ((x-3) x+3) y-3)\right.\\
\nonumber& &\left.+3)-1) \left(x y \left(25 x y \left(3 c_p+10\right)+28 \left(c_p+13\right)\right)-63\right)-42 m_s^2 (y (((x-2) x+2) y-2)+1) \left(2 x y \left(c_p+6\right)-15\right)\right)\\
\nonumber& &-7 s x y m_c^2 m_s (x y-1)^2 (y (((x-3) x+3) y+x-3)+1) \left(x y \left(11 x y \left(c_p+6\right)+4 c_p+99\right)-25\right)-14 m_c^3 m_s^2 (y (y (x\\
\nonumber& &\left.((x-3) x+3) y-3)+3)-1) \left(x y \left(x y \left(c_p+6\right)-15\right)+10\right)\right)+(x-1) x (y (y (x ((x-3) x+3) y-3)+3)-1)\\
\nonumber& &F(s,x,y)^3 \left(21 m_s \left(x y \left(4 x y \left(c_p+6\right)+2 c_p+57\right)-35\right)+2 x y m_c \left(-10 x^2 y^2 \left(3 c_p+10\right)-28 x y \left(c_p+8\right)\right.\right.\\
\nonumber& &\left.\left.+7 \left(c_p+18\right)\right)\right)+21 F(s,x,y)^2 \left(2 (x-1) (y-1) m_c \left(3 m_s^2 (y (((x-2) x+2) y-2)+1) \Big(x y \left(x y \left(c_p+6\right)-15\right)\right.\right.\\
\nonumber& &\left.+10\Big)-s x^2 y^2 (y (y (x ((x-3) x+3) y-3)+3)-1) \left(5 x^2 y^2 \left(3 c_p+10\right)+x y \left(11 c_p+78\right)-2 \left(c_p+12\right)\right)\right)\\
\nonumber& &+x m_c^2 m_s (x y-1)^2 (y (((x-3) x+3) y+x-3)+1) \left(x y \left(4 x y \left(c_p+6\right)+2 c_p+57\right)-35\right)+3 s (x-1) x (y-1) y\\
\nonumber& &\left. m_s (y (y (x ((x-3) x+3) y-3)+3)-1) \left(x y \left(11 x y \left(c_p+6\right)+4 c_p+99\right)-25\right)\right)+6 s x (y-1)^2 y^2 m_c (y (y (x ((x\\
\nonumber& &-3) x+3) y-3)+3)-1) \left(7 s x y m_c m_s (x y-1) \left(2 x y \left(c_p+6\right)+15\right)+14 m_c^2 m_s^2 \left(2 x y \left(c_p+6\right)-15\right)-2 s^2\right.\\
\nonumber& &\left.\left.(x-1) x^2 (y-1) y^3 \left(x y \left(3 c_p+10\right)+14\right)\right)\right)\Bigg\},\\
\nonumber\rho^{\langle\bar{s}\sigma\cdot Gs\rangle}_{5,11;A(S)}(s)&=&\int^{x_{max}}_{0}dx\int^{y_{max}}_{y_{min}}dy\Bigg\{\frac{c_1 x^2}{1536 \pi ^3 (y-1)^3} m_c (x y-1) F(s,x,y) \Big(3 s (y-1) y (x y(55 x y+56)-9) F(s,x,y)+4 (x y\\
\nonumber& & (5 x y+8)-3) F(s,x,y)^2+12 s^2 x (y-1)^2 y^3 (5 x y+4)\Big)+\frac{c_2 x^3}{1024 \pi ^3 (x-1) (y-1)^4} m_c (y (((x-2) x+2) y-2)\\
\nonumber& &+1) F(s,x,y) \Big(3 s (y-1) y (x y (11 x y+14)-3) F(s,x,y)+4 (x y (x y+2)-1) F(s,x,y)^2+12 s^2 x (y-1)^2 y^3\\
\nonumber& & (x y+1)\Big)\Bigg\},\\
\nonumber\rho^{m_s\langle\bar{s}\sigma\cdot Gs\rangle}_{5,11;A(S)}(s)&=&\int^{x_{max}}_{0}dx\int^{y_{max}}_{y_{min}}dy\Bigg\{-\frac{c_1 x}{1152 \pi ^3 (y-1)^2} \left(-3 s x (y-1) y^2 F(s,x,y) \left(6 m_c^2 (4 x y-3)+s (x-1) (y-1) y \left(x y \left(-4 c_p\right.\right.\right.\right.\\
\nonumber& &+125 x y+104\Big)-9\Big)\Big)-3 F(s,x,y)^2 \left(3 m_c^2 (x y-1) (2 x y-1)+s (x-1) x (y-1) y^2 \left(c_p (2-11 x y)+x y (175 x y\right.\right.\\
\nonumber& &\left.\left.+156)-24\right)\right)-(x-1) x y \left(-4 x y c_p+c_p+2 x y (25 x y+32)-18\right) F(s,x,y)^3-6 s^3 (x-1) x^2 (y-1)^3 y^5 (5 x y\\
\nonumber& &+4)\Big)+\frac{c_2 x^2 (x y-1)}{1024 \pi ^3 (x-1) (y-1)^3} \left(3 s x (y-1) y^2 F(s,x,y) \left(4 m_c^2 (x y-1)+s (x-1) (y-1) y (x y (25 x y+26)-3)\right)\right.\\
\nonumber& &+3 F(s,x,y)^2 \left(m_c^2 (x y-1)^2+s (x-1) x (y-1) y^2 (x y (35 x y+39)-8)\right)+2 (x-1) x y (x y (5 x y+8)-3)\\
\nonumber& &\left.F(s,x,y)^3+6 s^3 (x-1) x^2 (y-1)^3 y^5 (x y+1)\right)\Bigg\},\\
\nonumber\rho^{\langle\bar{s}s\bar{s}s\rangle}_{5,11;A(S)}(s)&=&0,\\
\nonumber\rho^{\langle\bar{s}s\rangle\langle\bar{s}\sigma\cdot Gs\rangle}_{5,11;A(S)}(s)&=&0,\\
\nonumber\rho^{\langle g_sG^2\rangle\langle \bar{s}s\rangle}_{5,11;A(S)}(s)&=&\int^{x_{max}}_{0}dx\int^{y_{max}}_{y_{min}}dy\Bigg\{\frac{c_1 x^3}{4608 \pi ^3 (x-1)^3 (y-1)^4} m_c (y (y (x ((x-3) x+3) y-3)+3)-1) \left(F(s,x,y) \left(4 m_c^2 (x y (x y\right.\right.\\
\nonumber& &\left.+2)-1) (x y-1)+3 s (x-1) (y-1) y (x y (11 x y+14)-3)\right)+6 (x-1) (x y (x y+2)-1) F(s,x,y)^2+s (y-1) y\\
\nonumber& &\left.\left(m_c^2 (x y (11 x y+14)-3) (x y-1)+6 s (x-1) x (y-1) y^2 (x y+1)\right)\right)\Bigg\}+\int^{1}_{0}dx\int^{1}_{0}dy\frac{c_1 s^2 x^4 y^3}{2304 \pi ^3 (x-1)^3 (y-1)^2}\\
\nonumber& & m_c^3 \left(x^2 y^2-1\right) (y (y (x ((x-3) x+3) y-3)+3)-1),\\
\nonumber\rho^{m_s\langle g_sG^2\rangle\langle \bar{s}s\rangle}_{5,11;A(S)}(s)&=&\int^{x_{max}}_{0}dx\int^{y_{max}}_{y_{min}}dy\Bigg\{\frac{c_1 x^2}{2304 \pi ^3 (x-1)^3 (y-1)^3} m_c^2 (x y-1) \left((x-1) \Big(y \Big(x \Big(y \Big(x \Big(y \Big(x \Big(10 ((x-3) x+3) y^2+26 (x\right.\\
\nonumber& &-3) y+23\Big)+48 y-36\Big)+7\Big)-12 y+18\Big)-3\Big)-6 y+6\Big)-3\Big) F(s,x,y)+m_c^2 (x y-1)^2 (y (((x-3) x+3) y+x\\
\nonumber& &-3)+1)+s (x-1) x (y-1) y^2 \left(x y \left(y \left(35 ((x-3) x+3) x y^2+(74 (x-3) x+117) y+72 x-105\right)+31\right)-12 (y\right.\\
\nonumber& &\left.-1) y-2\right)\Big)\Bigg\}+\int^{1}_{0}dx\int^{1}_{0}dy\Bigg\{\frac{c_1 s x^3 y^2}{4608 \pi ^3 (x-1)^3 (y-1)^2} m_c^2 (x y-1) (y (((x-3) x+3) y+x-3)+1) \Big(4 m_c^2 (x y\\
\nonumber& &-1)+s (x-1) (y-1) y (x y (25 x y+26)-3)\Big)\Bigg\},\\
\nonumber\rho^{\langle g_sG^2\rangle^2}_{5,11;A(S)}(s)&=&\int^{x_{max}}_{0}dx\int^{y_{max}}_{y_{min}}dy\Bigg\{\frac{c_1 x^5 y^4}{46448640 \pi ^5 (x-1)^2 (y-1)^2}m_c^4 \left(c_p (2 x y (15 x y+14)-7)+4 x y (25 x y+56)-126\right)\Bigg\}\\
\nonumber& &+\int^{1}_{0}dx\int^{1}_{0}dy\Bigg\{\frac{c_1 s x^5 y^5}{6635520 \pi ^5 (x-1)^2 (y-1)} m_c^4 \left(c_p (x y (15 x y+11)-2)+2 x y (25 x y+39)-24\right)\Bigg\},\\
\nonumber\rho^{\langle g_sG^2\rangle\langle \bar{s}\sigma\cdot Gs\rangle}_{5,11;A(S)}(s)&=&-\int^{x_{max}}_{0}dx\int^{y_{max}}_{y_{min}}dy\Bigg\{\frac{c_1 x^2}{18432 \pi ^3 (x-1)^3 (y-1)^3} m_c (y (y (x ((x-3) x+3) y-3)+3)-1) \Big(12 (x-1) (x y (5 x y\\
\nonumber& &+8)-3) F(s,x,y)+4 m_c^2 (x y (5 x y+8)-3) (x y-1)+3 s (x-1) (y-1) y (x y (55 x y+56)-9)\Big)\\
\nonumber& &+\frac{c_2 x^3 y^2}{3072 \pi ^3 (x-1)^2 (y-1)^2} m_c^3 (x y-1) (x y (x y+2)-1)\Bigg\}+\int^{1}_{0}dx\int^{1}_{0}dy\Bigg\{\frac{c_1 s x^2 y}{18432 \pi ^3 (x-1)^3 (y-1)^2} m_c (y (y (x\\
\nonumber& & ((x-3) x+3) y-3)+3)-1) \Big(m_c^2 (x y (55 x y+56)-9)(x y-1)+6 s (x-1) x (y-1) y^2 (5 x y+4)\Big)\\
\nonumber& &+\frac{c_2 s x^3 y^3}{12288 \pi ^3 (x-1)^2 (y-1)} m_c^3 (x y-1) (x y (11 x y+14)-3)\Bigg\},\\
\nonumber\rho^{m_s\langle g_sG^2\rangle\langle \bar{s}\sigma\cdot Gs\rangle}_{5,11;A(S)}(s)&=&\int^{x_{max}}_{0}dx\int^{y_{max}}_{y_{min}}dy\Bigg\{\frac{c_1 x}{13824 \pi ^3 (x-1)^2 (y-1)^2} m_c^2 (x y-1) \left(y \left(-x c_p (4 x y-1) (y (((x-3) x+3) y+x-3)+1)\right.\right.\\
\nonumber& &\left.\left.+y \left(x \left(x \left(50 ((x-3) x+3) x y^3+6 (19 (x-3) x+32) y^2+6 (19 x-29) y+37\right)-18 (y-2)\right)-18\right)+18\right)-9\right)\\
\nonumber& &+\frac{c_2 x^2 y }{12288 \pi ^3 (x-1)^2 (y-1)^2}m_c^2 (x y (x y (x y (2 y (5 ((x-2) x+2) y+8 x-26)+7)+4 (8 y-5) y+7)-12 (y-1) y\\
\nonumber& &+3)-3)\Bigg\}+\int^{1}_{0}dx\int^{1}_{0}dy\Bigg\{\frac{c_1 x}{13824 \pi ^3 (x-1)^3 (y-1)^2} m_c^2 \left(3 m_c^2 (x y-1)^2 (2 x y-1) (y (((x-3) x+3) y+x-3)\right.\\
\nonumber& &+1)+s (x-1) x (y-1) y^2 \Big(-c_p (x y-1) (11 x y-2) (y (((x-3) x+3) y+x-3)+1)+x y \Big(y \Big(x \Big(y \Big(x \Big(175 ((x\\
\nonumber& &\left.-3) x+3) y^2+156 (x-3) y+12\Big)-57 y+525\Big)-202\Big)-468 y+450\Big)-120\Big)+18 (y-1) y-3\Big)\right)\\
\nonumber& &+\frac{c_2 x^2 y}{12288 \pi ^3 (x-1)^3 (y-1)^2} m_c^2 \Big(m_c^2 (x y-1)^2 (y (((x-2) x+2) y-2)+1)+s (x-1) x (y-1) y^2 (x y (y (x (y (35 ((x\\
\nonumber& &-2) x+2) y+39 x-148)+33)+78 y-62)+27)-16 (y-1) y-2)\Big)\Bigg\},\\
\end{eqnarray}
}
where the coefficient $c_p=1$ for current $J_{5,\mu\nu}^{A(S)}$  while $c_p=-1$ for current $J_{11,\mu\nu}^{A(S)}$.

\end{document}